\documentclass[preprint]{aastex}
\usepackage{lscape}
\usepackage{rotating}
\usepackage{amsmath}
\usepackage{amssymb}
\usepackage{amsfonts}
\usepackage{epsfig}
\usepackage{natbib}
\input epsf
\def\1{\'\i}

\begin{document}

\title{Probing the Dark Flow signal in WMAP 9 yr and PLANCK 
cosmic microwave background maps.}

\author{
F. Atrio-Barandela\altaffilmark{1}, 
A. Kashlinsky\altaffilmark{2}, 
H. Ebeling\altaffilmark{3},
D.~J.~Fixsen\altaffilmark{4},
D. Kocevski\altaffilmark{5}
}
\altaffiltext{1}{F\1sica Te\'orica, Universidad de Salamanca, 37008 Salamanca, Spain;
email:atrio@usal.es}
\altaffiltext{2}{NASA Goddard Space Flight Center and SSAI, Observational Cosmology Lab,
Greenbelt, MD 20771 USA; email:Alexander.Kashlinsky@nasa.gov}
\altaffiltext{3}{Institute for Astronomy, University of Hawaii,
Honolulu, HI 96822 USA; email:ebeling@ifa.hawaii.edu}
\altaffiltext{4}{NASA Goddard Space Flight Center and UMCP, 
Observational Cosmology Lab, Greenbelt, MD 20771 USA; email:Dale.Fixsen@nasa.gov}
\altaffiltext{5}{Chemistry-Physics Building, University of Kentucky,
Lexington, KY 40508 USA; email:kocevski@pa.uky.edu}
%\altaffiltext{5}{Physics and Astronomy, 5800 Mayflower Hill Waterville, Maine 04901
%USA; email:dale.kocevski@colby.edu}

\begin{abstract}
The ``dark flow'' dipole is a statistically significant dipole found at the position 
of galaxy clusters in filtered maps of Cosmic Microwave Background (CMB) temperature  
anisotropies. The dipole measured in {\it WMAP} 3, 5 and 7 yr data releases was 1) mutually consistent, 2) roughly 
aligned with the all-sky CMB dipole and 3) correlated with clusters' X-ray luminosity.
We analyzed {\it WMAP} 9 yr and the 1st yr {\it Planck} data releases using 
a catalog of 980 clusters outside the Kp0 mask to test our earlier findings.
The dipoles measured on these new data sets are fully compatible with our
earlier estimates, being similar in amplitude and direction to our previous results
and in disagreement with the results of an earlier study by the {\it Planck} Collaboration. 
Further, in {\it Planck} datasets dipoles are found independent of frequency, ruling out the Thermal
Sunyaev-Zeldovich as the source of the effect. 
 Both, in {\it WMAP} and {\it Planck}, we find a clear correlation 
between the dipole measured at the cluster location in filtered maps with
the average anisotropy on the original maps, further proving that the dipole
is associated with clusters. The dipole signal is dominated
by the most massive clusters, with a statistical significance better than 99\%, 
slightly larger than in {\it WMAP}.  Since both data sets differ in foreground
contributions, instrumental noise and other systematics, the agreement
between {\it WMAP} and {\it Planck} dipoles argues against them being due 
to systematic effects in either of the experiments.
\end{abstract}


\maketitle

\section{Introduction.\label{sec:intro}}

Measurements of peculiar velocities with galaxies rely on distance 
indicators to subtract the Hubble expansion, and have achieved notable success  
in probing peculiar velocities using galaxy surveys out to $\le  100h^{-1}$Mpc 
(e.g. see review by Strauss \& Willick, 1995). However, individual galaxy distance 
indicator surveys are typically restricted to isolated parts of the sky, which 
should then be corrected for when reconstructing galaxy flow characteristics, 
such as the amplitude of the bulk flow on a given scale (e.g.  Watkins et al. 2009). 
SNIa surveys input highly accurately measured individual distances (Turnbull 
et al. 2012), but are sparse and require corrections for sky coverage effects (Rathaus 
et al. 2013). More critically, in all of such measurements the galaxy velocity is probed with
respect to the frame of the Hubble expansion and translated into the Cosmic Microwave
Background (CMB) rest frame after assuming that the {\it entire} CMB dipole is of 
purely kinematic origin caused by the Doppler effect due to the local motion of 
our Galaxy, Local Group etc. (Kogut et al. 1993, but see the 
discussion of Wiltshire et al.  2013). 

The kinematic Sunyaev-Zel'dovich (KSZ; Sunyaev \& Zel'dovich, 1972) measures
directly the peculiar velocity of clusters with respect to the CMB and does 
not require subtracting the velocity
due to the Hubble expansion. Therefore KSZ offers an alternative method
to probe peculiar velocity field at larger distances potentially inaccessible to 
galaxy distance indicator methodologies. Its main disadvantage
is that the temperature fluctuations due to the peculiar motion of individual clusters
are much smaller than the cosmological CMB signal, foreground emissions,
instrumental noise or the Thermal Sunyaev-Zeldovich anisotropies
(TSZ; Sunyaev \& Zel'dovich, 1970) from the
thermal motion of electrons in the potential well of clusters.
As a result, the peculiar velocity of a single cluster has yet to be determined.
Kashlinsky \& Atrio-Barandela (2000, hereafter KA-B) thus proposed 
a method to probe the bulk motion of clusters of galaxies collectively using all-sky 
CMB maps combined with an all-sky X-ray cluster catalog. They pointed out 
that one can construct a statistic, the dipole moment evaluated at cluster locations 
over a fixed aperture containing the entire X-ray emitting gas, which can probe the 
bulk flow down to cosmologically interesting levels for the 
{\it WMAP} and {\it Planck} instrumental configurations.
The KA-B method requires filtering out the primary CMB component {\it without removing the
KSZ signal}, and isolating the TSZ contribution to the measured dipole. For the former,
in KA-B we proposed a variant of the Wiener filter, designed to minimize the
contribution from primary CMB with the known mean power spectrum, whereas the 
TSZ component can be attenuated if in clusters 
the gas X-ray temperature, $T_X$, decreases toward the outer parts as 
it was indeed found empirically (Atrio-Barandela et al. 2008).

The KA-B method was first applied to the 3-yr {\it WMAP} CMB data coupled with 
an extended cluster catalog where, surprisingly, a statistically significant dipole 
over the cluster apertures containing {\it zero} monopole was found 
for a volume of median/mean depth of $\sim 300h^{-1}$Mpc (Kashlinsky et al. 2008, 2009,
hereafter KABKE, KABKE2). Within the statistical and systematic 
calibration uncertainties this corresponded to the cluster sample 
moving at $\sim 600-1,000$km/s in the direction of the CMB dipole. KABKE
termed this the ``dark flow" speculating that it may be reflective of the 
{\it effective} motion across the entire cosmological horizon. If true, 
this is equivalent to at least a part of the all-sky CMB dipole being of 
primordial origin, a possibility that requires an isocurvature component in 
the primordial density field (Matzner 1980, Turner 1991, Mersini-Houghton \& Holman 2009).
Using a further expanded cluster catalog and {\it WMAP} 5-yr CMB maps 
Kashlinsky et al. (2010, hereafter KAEEK) showed that the cluster dipole correlated
with cluster properties, increasing in amplitude for the most X-ray luminous 
and massive clusters, as expected from SZ contributions (the TSZ contribution 
being small over the final apertures as evidenced by the zero monopole there). 
Atrio-Barandela et al. (2010, hereafter AKEKE) have developed - analytically 
and numerically - the formalism to understand the error budget of the KA-B method, 
which can and should be applied to any such measurement as a consistency 
check\footnote{As we will discuss later and as was pointed out numerous times, Keisler (2009) claims errors which violate 
the AKEKE analytical and numerical evaluations and are indicative of an error 
which he confirmed in private correspondence.}. Kashlinsky et al. (2011, hereafter KAE11) 
have shown that the results can be probed with public cluster data which they
have posted for interested investigators at
\url{www.kashlinsky.info/bulkflows/data\_public}. The methodology of 
the analysis, the results and their potential implications  have 
been extensively reviewed in Kashlinsky et al. (2012, hereafter KAE12).

Motivated by the final {\it WMAP} 9 yr and {\it Planck} 1 yr data 
releases we have scrutinized our previous ``dark flow" measurements 
with the further developed methodology and present the results here. 
We do not address our interpretation of the signal here: throughout 
we refer as the ``dark flow signal" to the statistically significant dipole
remaining at cluster positions and with an amplitude which correlates 
with X-ray cluster luminosity pointing, within the uncertainties, in 
the direction of the all-sky CMB dipole. Because {\it WMAP} does 
not have the frequency coverage required to distinguish a KSZ dipole from 
the dipole generated by a random distribution of the TSZ anisotropy, we 
evaluated the final dipoles at apertures containing zero monopole.
Since the mean TSZ monopole is an upper bound on the TSZ generated
dipole, this aperture guarantees that the measured dipole was not due to
the TSZ effect. Importantly, Planck has measured on both
sides of the zero-TSZ frequency at 217 GHz and has provided the appropriate 
data to test whether the dipole contains a significant TSZ contribution. 
For this data we just require the aperture to be large enough for the 
errors integrate down and leave a statistically significant dipole. We will 
show that, at the same aperture, the {\it Planck}-based results are fully consistent
with those of {\it WMAP},  providing a very important consistency check.
The measured dipole turns out to be independent of frequency and is consistent 
with the CMB black body energy spectrum and therefore, cannot be due to TSZ 
or foreground residuals since those components vary with frequency.

We find the same results as before with the {\it WMAP} 9 yr data analysis, 
but given the lower noise levels of that dataset and the new methodology 
here we can isolate the signal better. We then apply the methodology to 
{\it Planck} 1 yr data and we find full consistency with the {\it WMAP} 
results. There appears a statistically significant ``dark flow'' signal at 
cluster locations with the dipole amplitude which correlates with
cluster X-ray luminosity and the direction pointing within the uncertainties 
to the direction of the all-sky CMB dipole. If the measured signal with all 
its properties can arise from something other that KSZ, we would welcome this 
discussion. 

This paper is structured as follows: For completeness we briefly revisit 
the methodology, the data processing pipeline and the error budget of the 
KA-B measurement. Then in Sec. 3 we present the analysis of the 
{\it WMAP} 9 yr data, which supports empirically the error budget 
estimations derived in Sec. 2 and our previous measurements. Sec. 4 
addresses our measurement of the dark flow signal with the {\it Planck} 
1 yr data. We find full consistency between the {\it WMAP} and {\it Planck} 
results, except that for the map at 30~GHz and in particular cluster 
configurations, that could be  affected by low-level systematics, consistent 
with the effects of striping due to the {\it Planck} observing strategy. The
``dark flow'' measured in {\it Planck} is significant at better than the 
99\% confidence level. When combined with the fact that the signal correlates 
with cluster X-ray luminosity and points in the direction of the all-sky 
CMB dipole, the significance of the existence of the primordial contribution 
to the CMB dipole, known as "Dark Flow", is even larger.
Throughout this paper we use the X-ray cluster catalog compiled for the 
KAEEK study. A more advanced and expanded catalog is now being worked on 
and upon its completion we will present the results from its application.

\section{Methodology, Data Processing and Error Budget.}
\label{sec:filter}

\subsection{KA-B method}

A cluster in the direction $\hat{n}$, moving with a peculiar velocity 
$\vec{v}$, will generate a temperature anisotropy 
$\Delta T_{\rm KSZ}=-T_0\tau (\vec{v}\cdot\hat{n}/c)$, where
$\tau$ is the projected electron density along the line of sight,
$c$ the speed of light and $T_0$ is the CMB blackbody temperature. 
A sample of clusters randomly located in the sky moving with an average 
velocity $\vec{V}_{\rm bulk}$ will produce a temperature anisotropy
$\Delta T_{\rm KSZ}=- T_0\tau (V_{\rm bulk}/c)\cos\theta$, where $\theta$
is the angle with respect to the apex of the motion. 
At the position of clusters, microwave temperature anisotropies 
have several components: primary CMB, TSZ and KSZ components, foreground
residuals and instrument noise.  KA-B estimated how these terms integrated 
down with many clusters concluding that at the resolution of {\it WMAP} and {\it Planck} 
channels the dominant contribution to the noise of the KSZ measurement would be
from primary CMB anisotropies. KA-B proposed to use the known statistical properties
of the primary CMB to filter out this contribution and increase 
the signal-to-noise of the probed KSZ term.
The KA-B proposed filter minimizes the difference $\langle(\Delta T-{\cal N})^2\rangle$,
with ${\cal N}$ being the instrumental noise (Kashlinsky et al. 2009). 
AKEKE have shown analytically and numerically that it effectively removes
the primary CMB signal down to the cosmic variance. In $\ell$-space the KA-B filter is
$F_\ell=(C_\ell^{\rm sky}-C_\ell^{\rm th}B_\ell^2)/C_\ell^{\rm sky}$
where $C_\ell^{\rm sky}$ is the actual realization of the radiation power
spectrum in {\it our sky} that includes noise, TSZ, KSZ, foreground residuals,
and primary CMB; $C_\ell^{\rm th}$ is the power spectrum
of the $\Lambda$CDM model  realization that best fits the data, 
and $B_\ell$ is the antenna beam. 

\subsection{Data processing pipeline}

{\it WMAP} and {\it Planck} have measured the microwave sky at different frequencies with 
varying angular resolution. We implement our filter taking into account the specifics of 
each of the {\it WMAP} Differencing Assembly (DA) or the {\it Planck} channels.
Our pipeline for measuring the dark flow signal works as follows:

\begin{enumerate}
\item We start with foreground-cleaned all-sky microwave maps.
\item{} The data of each channel are multiplied by the Galactic and point
source mask. To facilitate comparison with our previous results we chose
the {\it WMAP} Kp0 mask. 
\item Next, monopole, dipole, and quadrupole are subtracted
from the regions outside the mask. Then we compute the multipole expansion
coefficients $a_{\ell m}$ correcting for the mask.
\item{} The $a_{\ell m}$ coefficients are multiplied by the filter $F_\ell$ before
transforming them back into real space to create the filtered map.
Since the quadrupole and octopole are
aligned with the dipole the filter is set to zero for $\ell\le 3$ to avoid any 
cross-talk between those scales that could mimic a dipole. 
\item{} The monopole and dipole outside the mask are removed from
the filtered maps.
\item{} The dipoles are computed at the cluster positions using fixed 
aperture for all clusters for a given depth, X-ray luminosity cutoff configuration. 
\end{enumerate}

Our first results, presented in Atrio-Barandela et al. (2008) and KABKE1,2
were obtained using different apertures. We estimated the 
size of the region that emitted 99\% of the X-ray flux, $\theta_X$, for
each cluster and computed dipoles in units of $\theta_X$. The results were 
found to be very similar to those using a fixed aperture zero monopole for 
all clusters so we present the results using fixed apertures where errors 
are simpler to compute and can be evaluated by analytic means 
providing multiple cross-checks (AKEKE).

As indicated in the introduction, KAEEK showed that when binned by 
cluster X-ray luminosity, the cluster dipole measured in filtered maps correlated 
with central TSZ anisotropy in unfiltered maps, with larger amplitudes corresponding to 
the most X-ray luminous clusters, as expected from SZ contributions.
Due to the inhomogeneous distribution of clusters on the sky, the 
mean TSZ anisotropy (or monopole) could generate a significant dipole and/or
other higher order multipoles. It is important to demonstrate that the 
measured dipole was not due to the TSZ effect. {\it WMAP} operated in 
the Rayleigh-Jeans CMB regime and did not provide enough direct  information to 
subtract the TSZ contributions from the measured dipoles.
To  ensure that in {\it WMAP} data the measured dipole was not dominated by 
the TSZ monopole we used the fact that all TSZ multipoles due to the 
inhomogeneous distribution of clusters on the sky, including the dipole, 
would be bounded from above by the monopole. Then, in step [6] we repeated 
the measurement for different apertures and selected the dipole measured 
at the {\it zero monopole} aperture to ensure that the TSZ component did 
not contribute to the measurement. This aperture is no longer necessary 
when using {\it Planck} data, where the TSZ vanishes at 217GHz and any dipole 
there will be free from TSZ contributions.

\subsection{Error budget}

We compute errors numerically using the same configurations and apertures 
that were used to evaluate the dipole at the cluster positions in KAEEK. 
In KAE12 we discussed four 
different methods to evaluate numerically the errors and showed their mutual consistency
(Sec. 10.3). Our dipole is measured at cluster pixels so the error on this 
measurement is determined by the distribution of the random dipoles in 
the data away from the actual clusters. We evaluate these random dipoles 
by placing filled aperture discs at random positions in the sky with the same 
angular extent as was used to measure the dipole at the cluster location. 
We remove all pixels within 80 arcmin from the center of all known clusters 
to make sure that the randomly distributed discs do not overlap with them.
By using the realization of the primary CMB as given by the actual sky to 
measure the dipole and its error we take into account the effect of
all possible systematics existing in the filtered data, such as foreground
residuals, inhomogeneous and correlated instrument noise, as well as any 
artifact that could have been introduced by our pipeline like mode coupling and 
power leakage between the galactic mask and the cosmological signal remaining in 
the filtered data. In Atrio-Barandela (2013) we discussed the different biases 
and inefficiencies that exist between different types of simulations that can 
result on overestimating the errors. In AKEKE we developed an analytical insight  
to detail the different contributions to the error bars and their properties
when the instrumental noise is Gaussian-distributed and foreground
residuals are negligible. More details are given in AKEKE and KAE12 (Sec. 10.3), where 
it is shown that numerical simulations of the actual CMB sky give errors 
in excellent agreement with the analytical theory. This formalism clarifies 
the relation between the errors of the monopole and of the three dipole 
components and their scaling with the number of clusters; it is briefly 
summarized below.

The filtered maps have variance $\sigma_{fil}^2=(1/4\pi)
\sum(2\ell+1)F_\ell^2C_\ell^{\rm sky}$.
While the filter erases a large fraction of the primary CMB anisotropy, {\it it 
leaves a residual due to cosmic variance that is common to all frequencies}.
The realization of the radiation power spectrum as seen from our
location, $C_\ell^{\rm sky}$, differs from the underlying power spectrum
$C_\ell^{th}$ by a random variable of zero mean and (cosmic) variance 
$\Delta_\ell=(\ell+\frac{1}{2}) C_\ell^{th}/f_{sky}$ (Abbott \& Wise 1984). 
In addition, the instrument noise is also present with a power 
spectrum $N_\ell$. Neglecting foreground residuals in the foreground-cleaned 
maps, AKEKE have demonstrated that the variance of the filtered map is given by 
propagating the cosmic variance (also, see Sec 10.3.1 of KAE12 for a more 
detailed derivation) 
\begin{equation}
\sigma_{fil}^2=\frac{1}{4\pi}\sum(2\ell+1)
\left[\frac{\Delta_\ell}{C_\ell^{th}+\Delta_\ell+{\cal N}_\ell}+
\frac{{\cal N}_\ell}{C_\ell^{th}+\Delta_\ell+{\cal N}_\ell}\right]=
\sigma_{CV,fil}^2+\sigma_{{\cal N},fil}^2(t_{obs})
\label{eq:sigma_fil}
\end{equation}
This expression is valid in the limit of zero TSZ contribution 
and so does not reflect the fact that the final errors will depend on 
the radius of the aperture chosen around each cluster, so they need to 
be computed numerically. An error-aperture dependence is to be expected since
the residual CMB and noise have different spectra as shown in AKEKE 
(see Fig.~1); the residual cosmological CMB signal dominated at
$\ell\le 300$ while the noise dominated at $\ell\ge 300$. By taking
larger apertures the instrument noise  integrates down and the residual
CMB dominates the error budget. For this reason 
the final errors on {\it WMAP} and {\it Planck} will be similar even if
they have very different noise levels. For {\it WMAP} we chose an
aperture that guarantees that there are no contributions to the dipole due to the
TSZ effect. At this aperture errors integrate down and leave a 
statistically significant dipole.
Nevertheless, eq.~(\ref{eq:sigma_fil}) is very 
instructive since it shows that the variance of the filtered maps 
depends  mainly on two components: (a) the residual CMB not removed 
by the filter due to cosmic variance (CV) and (b) the noise that decreases 
with increasing time of observation $t_{obs}$. Fig.~1 of AKEKE demonstrates 
empirically the high accuracy of the above expression.

By construction, the filtered maps have no intrinsic monopole or dipole.
Since we measure these two moments from a small fraction of the sky, 
our limited sampling generates an error due to the (random) distribution of 
these quantities around their mean zero value. This error is proportional to the
rms dispersion of the filtered map, the size of the fixed aperture
around each cluster and the number of clusters $N_{cl}$ of the
catalog. For a fixed aperture, the cosmic variance term of
eq~(\ref{eq:sigma_fil}) scales as $N_{cl}^{-1}$ while 
the noise term scales as $(N_{cl}N_{pix})^{-1}$, being $N_{pix}$
the number of pixels within a fixed aperture around each cluster.
The sampling variances of the monopole and three components of 
the dipole $(a_0,a_{1m})$ depend on how homogeneously clusters sample the
sky. A direct calculation shows that the monopole ($\sigma_0$) and 
three dipole ($\sigma_m$) errors are:
\begin{equation}
\sigma_0\equiv\langle a_0^2\rangle^{1/2}\approx\frac{\sigma_{fil}}{\sqrt{N_{cl}}},
\qquad \sigma_m\equiv \langle a_{1m}^2\rangle^{1/2}=
\frac{\sigma_0}{\langle\hat{n}_i^2\rangle}
\label{eq:error}
\end{equation}
where $\hat{n}_i$ are the clusters' direction cosines. If clusters were
homogeneously distributed in the sky  then $\langle\hat{n}_i^2\rangle=1/3$
and the errors on the dipoles would be related to that on the monopole as
expected: $\sigma_m=\sqrt{3}\sigma_0$, since three quantities are evaluated from
the same data as the monopole. As we will discuss in Sec.~\ref{sec:3.2},
this expression is only approximately true
since, due to the galactic mask, the error on the X and Y components 
of the dipole is slightly larger than that of the Z-component
(see AKEKE, Sec 10.3.2 and Fig 10.7 of KAE12, for a detailed discussion).

The analytical formalism summarized in this section neglects the 
contribution of possible foreground residuals but {\it they are already included in
the numerically computed statistical uncertainties} since, as mentioned 
above, we compute errors using the same realization of the sky than the 
data and, therefore, including foreground residuals as well as all 
other systematics. We find that eqs.~(\ref{eq:sigma_fil}) and
(\ref{eq:error}) agree with the errors found in simulations from the 
actual sky showing empirically that foreground contributions are small.

\subsection{Filtering and Noise}\label{subsec:filter-noise}

The lower noise levels of {\it Planck} as compared with {\it WMAP}
as well as different and independent systematics, allow a second 
and in some ways independent measurement of the KSZ signal, providing
further test of our filtering scheme. The KA-B filter has been designed 
to remove the cosmological signal by minimizing the difference between 
the data and the instrumental noise, i.e., the filter minimizes 
$\langle(\Delta T-{\cal N})^2\rangle$. The filter oscillates
around zero, $F_\ell\simeq 0$, where the noise is negligible and 
$F_\ell\simeq 1$ where the noise dominates. If the noise decreases, 
the filter will remove all signals down to the limit imposed by cosmic 
variance (KABKE2). Since cosmic variance decreases as $\ell^{-1/2}$, lowering 
the noise implies that the filter will remove the signal at high $\ell$'s 
more effectively.  Then, whether the TSZ and KSZ signals survive or not, 
our filtering will depend on how these contributions are distributed in 
$\ell$-space. For instance, the average TSZ anisotropy when evaluated 
at the cluster locations is a monopole and, in the absence of mask, 
filtering will distribute it preferentially to even multipoles while the  
bulk flow due to all the clusters is a dipole and it will be
distributed preferentially to odd multipoles. In other words,
the filter depends on the instrument noise (see Sec~2.1) and 
maps with noise of different amplitude will give different
filtered maps; this will change the redistribution
of the TSZ and KSZ components, intrinsic CMB and foreground
residuals, changing the amplitude and direction
of the measured dipole. Therefore, due to the difference in
noise amplitude and properties, it is important to compare the 
results obtained with {\it WMAP} and {\it Planck} to isolate the 
effect of systematics. 

\begin{figure*}
\centering
\includegraphics[width=18cm]{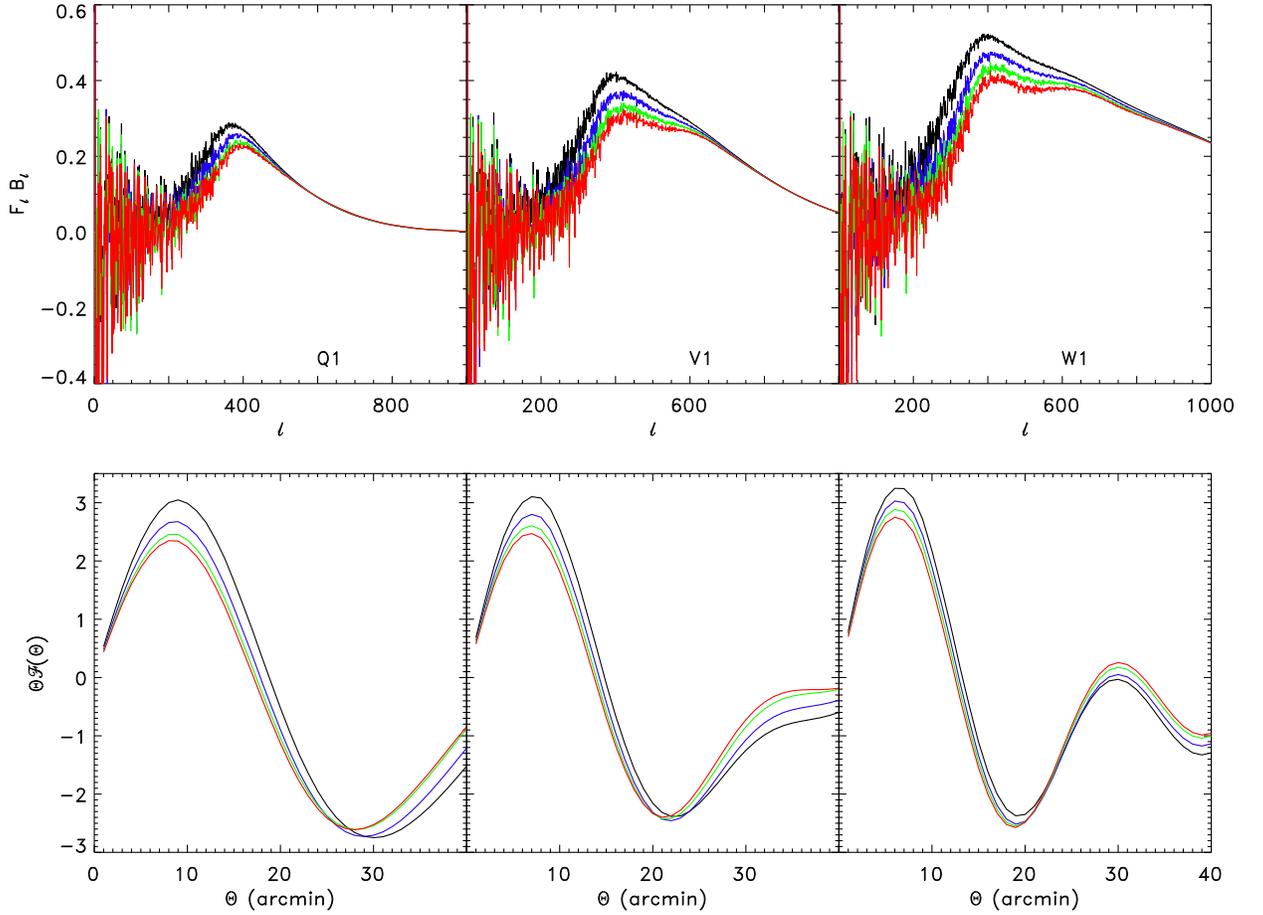}
\caption{\small Wiener filters constructed with the 3, 5, 7, 9 yr 
data (and noise levels). Top panels correspond to the filter in 
$\ell$-space and the bottom panels to the same filters in real space. In each
panel, black, blue, green and red curves from top to bottom correspond to
the 3 to 9 yr data. The galactic mask used was the {\it WMAP} Kp0 mask.
}
\label{fig:ak1}
\end{figure*}

\section{The dark flow dipole in WMAP 9 yr data.}
\label{sec:wmap}

We first present the results of our analysis of the dark-flow signal in the 
final {\it WMAP} 9 yr data. To facilitate the comparison with our earlier 
results, we use the Kp0 mask to remove the Galaxy and the X-ray cluster 
catalog assembled for the KAEEK study. This catalog contains 980 clusters 
outside {\it WMAP} Kp0 mask, with redshifts $z\le 0.25$ and X-ray 
luminosities in the ROSAT $(0.1-2.4){\rm KeV}$ band of $L_X\ge 0.2\times 
10^{44}$erg/s. Of those, 598 have X-ray luminosities $L_X>10^{44}$erg/s. 
We consider four cumulative redshift bins, selecting clusters by 
redshift: $z\le(0.12, 0.16, 0.2, 0.25)$. In each redshift bin, we define 
three independent cluster subsamples according to their luminosity. 
These subsamples are $L_X=(0.2-0.5,0.5-1.0,>1.0)$ (in the same units 
as before) for clusters with $z<0.12$ and $L_X=(0.5-1.0,1.0-2.0,>2.0)$ 
for all other bins. The number of clusters and other properties of each 
subsample are given in Table~1 of KAAEK. In total we only have 11 different 
bins since the bins $z<0.2$ and $z<0.25$ with $L_X<10^{44}$erg/s differ 
by two clusters and their results are almost identical.

\begin{figure*}
\centering
\includegraphics[width=18cm]{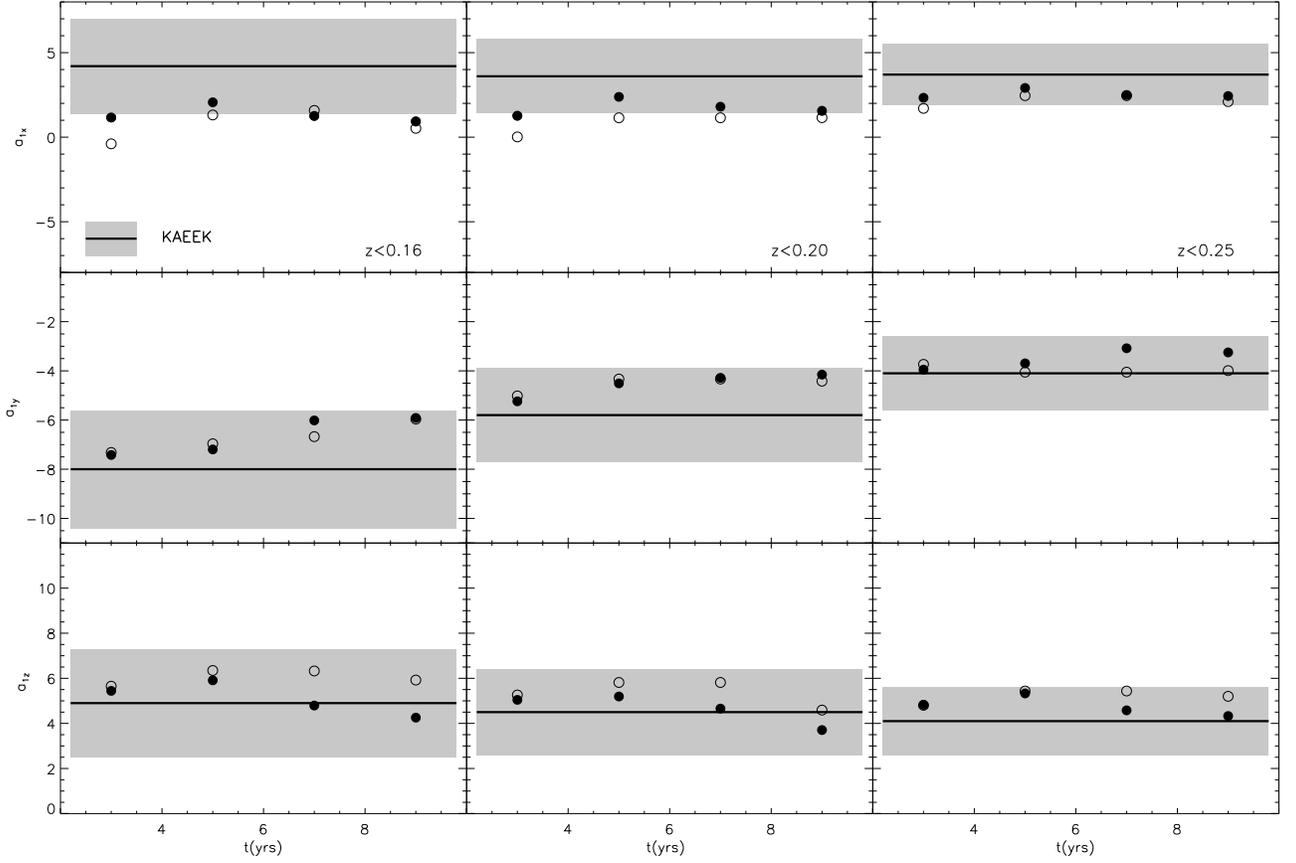}
\caption{\small Dipoles measured in {\it WMAP} 9 yr maps filtered with 
3, 5, 7 and 9 yr noise filters for subsamples with $L_X\geq 2\times 
10^{44}$erg/s and the indicated redshifts. These bins contain 130, 
208 and 322 clusters, respectively. Filled circles 
represent the averages over all 8 {\it WMAP} DA's; open circles are 
the averages over the 4 W-band DA's. The solid lines and shaded regions 
correspond to the KAEEK values and error bars.
}
\label{fig:ak2}
\end{figure*}

\subsection{WMAP filtering and results}

In KABKE, KAEEK and KAE11 we have analyzed the subsequent data releases 
of {\it WMAP} 3, 5 and 7 yr data, respectively. We have consistently
constructed the filter from the same data that we used to compute the
dipoles. With each release, the noise level in the map has decreased,
changing the filter. Motivated by the discussion in the 
Sec.~\ref{subsec:filter-noise} we can now test the robustness of the 
detected dipole signal with respect to the noise level of the filter; 
i.e., we can test the effect of the noise in redistributing the 
signal in the $\ell$-space of the filtered maps and its effects on the 
measured dipole. For this purpose, we have constructed four
filters for each of the eight single frequency all-sky CMB maps using
the data from the 3, 5, 7 and 9 yr releases. During these integrations 
the rms instrument noise has been reduced by a factor $\sqrt{3}$ and so each 
subsequent filter would remove progressively a larger fraction of the intrinsic
CMB signal. The four different filters were then applied to the {\it WMAP} 9 yr 
data of the ultimate noise achieved with that instrument. 
The filters in multipole (top panels) and angular (bottom panels)
space are shown in Fig.~\ref{fig:ak1}, where one can see also the differences in 
the maps of different resolution going from Q at $\sim 30'$ to W at $12'$. 
Since the noise is largest at the W DA's and lowest at Q more structure
survives in the former than in the latter filter. However, 
combining the four W DA's decreases the instrument noise by a factor of 2.

The overall S/N of the KAEEK measurement is driven by the most luminous 
X-ray clusters as should be if the dipole arises from the SZ cluster 
components. In Fig.~\ref{fig:ak2} we show the dipole at zero monopole aperture 
at the positions of the brightest clusters with $L_X\geq 2\times 10^{44}$erg/sec 
from Table 1 of KAEEK. The three panels correspond to different redshifts, 
as indicated. These bins contain 130, 208 and 322 clusters, respectively. 
Solid circles correspond to averaging over all eight {\it WMAP} DA's (Q, V, W bands) 
and open circles to averaging over four W DA's which can resolve clusters better. 
The  horizontal axis indicates the data used to construct the filter which is,
as we have indicated above, always applied to the the {\it WMAP} 9 yr data release. 
For the different filters, the zero monopole aperture changes from 20-30arcmin 
depending on the chosen cluster sample. Nevertheless, the difference with the
results at a fixed aperture, say 25 or 30arcmin, are negligible since the
residual monopole is always small, proving that the dipole is not
contaminated by the TSZ monopole. The black line and shaded area correspond to the 
dipoles measured in KAEEK, obtained from {\it WMAP} 5 yr data for
the same cluster samples. The figure shows a reassuring consistency between 
the 9 yr {\it WMAP} data (with any filter) with what was obtained in KAEEK 
for 5 yr {\it WMAP} data, which in turn have been demonstrated to be 
consistent with 3- and 7-yr WMAP CMB data in KAEEK and KAE11.

\subsection{The error budget for WMAP filtering}\label{sec:3.2}

In AKEKE, KAE12 we have discussed the proper methods to compute error bars 
and have addressed their relative merits and intrinsic biases. In Sec.~10.3.2 
of KAE12 we compared four alternative methods and showed that they all give 
similar uncertainties. As indicated in Sec.~2.3, we compute error bars 
by choosing random positions in the sky, outside the known clusters and the 
mask, and evaluating dipoles subtended by a given aperture around these 
centers, referred as Method 1 in KAE12. Each run was done with several 
apertures and with different number of clusters, in the range 
$100<N_{\rm clus}<600$. We compute the monopole and dipole at 
those $N_{\rm clus}$ random positions using the Healpix {\it remove\_dipole} 
routine. Our errors are the rms deviations of all those monopoles and dipoles, 
which coincide with the 68\% confidence level for these demonstrably 
gaussian distributions. 

\begin{figure*}
\centering
\includegraphics[width=18cm]{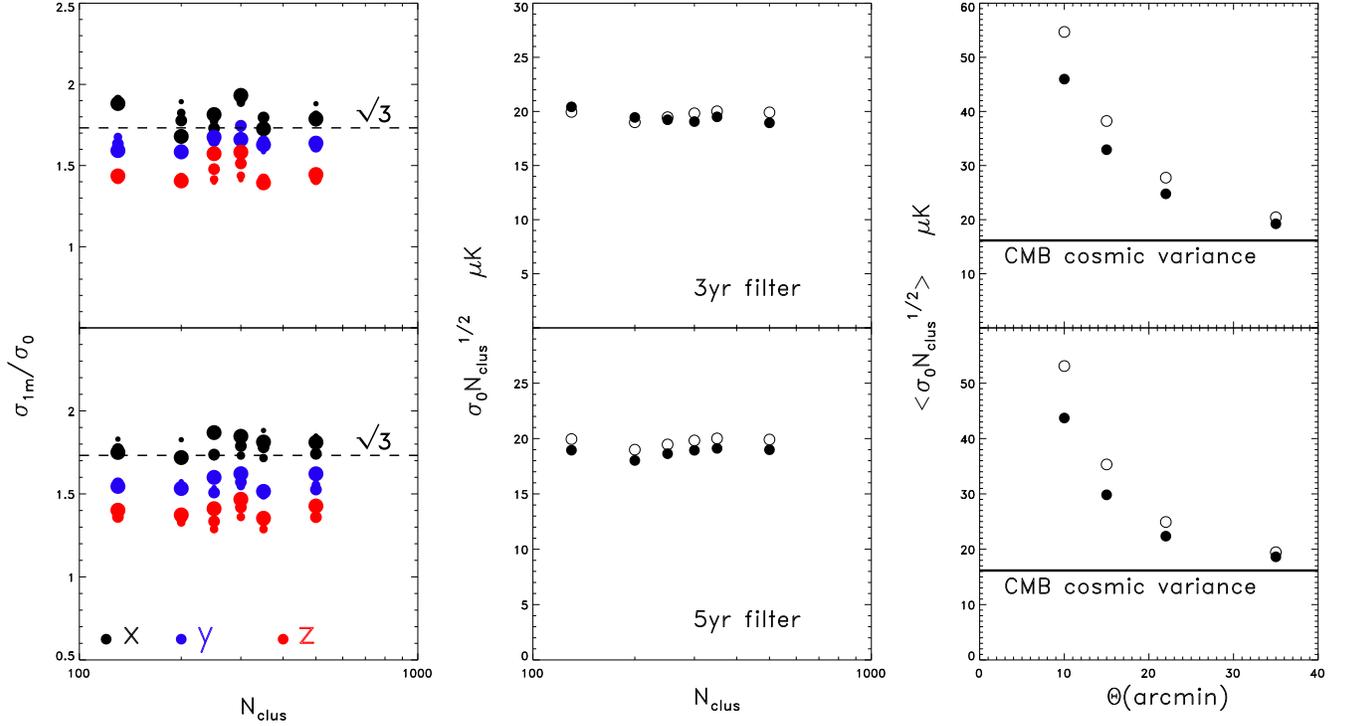}
\caption{\small Error estimates. In the left panels the black, blue, red 
circles correspond to the X, Y, Z components of the error. Circles of 
increasing size correspond to apertures of $10',15',22',35'$. The plot
show the robustness of the error estimates, driven predominantly by that 
of the monopole. The middle panels represent the monopole error vs the number of 
clusters for Method 1 (see Sec.~\ref{sec:3.2}).
Filled circles represent the average over all 8 DA's 
and open circles the average restricted to the 4 W-band DA's for apertures at
$\sim 30'$. The errors show good accuracy with the analytical theory developed in 
AKEKE and KAE12, Sec.~10.3.  In the right panel we show the error vs aperture
with the same convention than in the middle panels. The  thick horizontal line shows 
the zero noise cosmic variance limit $\sigma_{CV,fil}$ of the error 
(see eq~\ref{eq:sigma_fil}).}
\label{fig:ak3}
\end{figure*}

As discussed in AKEKE and above, the dipole error budget is driven almost 
entirely by the error on the monopole, $\sigma_0$, which should scale
as $\propto N_{\rm clus}^{-1/2}$. AKEKE have demonstrated, analytically 
and numerically, that the errors on the three dipole components should then be 
$\sigma_{1m}\simeq \sqrt{3}\sigma_0$ with $\sigma_{1x}>\sigma_{1y}>\sigma_{1z}$. 
The errors claimed by Keisler (2009) do not satisfy this and point to the 
flaw in his analysis namely, that his error budget is driven 
by the residual dipole outside the mask in the filtered map, which he failed to 
subtract prior to compute dipoles at random locations\footnote{It is unfortunate that, 
despite this, his claims are still occasionally cited at face value.},
as was demonstrated in AKEKE.  The left  panels of Fig. \ref{fig:ak3} show with 
simulations that for the {\it WMAP} 9 yr noise levels one obtains to good accuracy 
that $\sigma_{1x}\simeq 1.8\sigma_0, \sigma_{1y}\simeq 1.55\sigma_0, \sigma_{1z}
\simeq 1.4\sigma_0$ with a weak aperture dependence. {\it This confirms explicitly 
that the entire error budget is contained in $\sigma_0$.}

The middle panels of Fig. \ref{fig:ak3} show the results of simulated errors 
on $\sigma_0$ for various cluster configurations and the KABKE 3 and 5 yr 
filters for the $30'$ aperture which correspond approximately to the zero 
monopole aperture. Solid circles correspond to the average of all 8 DA's. The panels
demonstrate the accuracy of the scaling of $\sigma_0\propto N_{\rm clus}^{-1/2}$ or 
more explicitly $\sigma_0\simeq 20N_{\rm clus}^{-1/2}\mu$K with a weak dependence 
on the selected filter, which is valid in the limit of the instrument 
noise levels corresponding to 9-yr WMAP and Planck data. 
{\it This quantifies the errors explicitly}.

The right panels of Fig. \ref{fig:ak3} show the error on the monopole, 
$\sigma_0\times N_{\rm clus}^{1/2}$, as function of the aperture radius 
when averaging over all 8 DA's (filled circles) and 4 W-band DA's (open 
circles). This is compared with the component of eq.~\ref{eq:sigma_fil} 
resulting from the cosmic variance of primary CMB discussed in AKEKE, 
Sec. 10.3.1 of KAE12 and above, shown with a thick horizontal line. 
The three dipole components show a similar behavior, decreasing
with increasing aperture radius as the noise integrates down, and
are not shown. One can 
see explicitly that as the {\it WMAP} 9 yr instrument noise decreased with
increasing aperture, the monopole errors $\sigma_0$ approach this limit 
very accurately. Any filtering scheme should be able to evaluate similar 
expressions and then verify whether their particular claims are commensurate 
with this theoretically justified limit (cf. \citet{pir13}). {\it The final 
zero monopole aperture at $\simeq 30'$ (KABKE and KAEEK) is where the 
instrument noise (and foreground residuals) in {\it WMAP} 9 yr CMB maps 
contribute about $\sim 10\%$ to the total error, and so the latter 
is driven by cosmic variance from primary CMB}.

\subsection{The ``dark flow" dipole from the WMAP data}

In KAEEK we demonstrated that the measured dipole correlated with
cluster X-ray luminosity binning. We selected clusters by their redshift and we
divided the samples in three bins according to their X-ray luminosity,
$L_X$. We showed that clusters in the highest luminosity bin, with
$L_X[0.1-2.4KeV]\ge 2\times 10^{44}$erg/s, had larger monopoles and
larger dipoles than the other two bins. In fact, 
at the cluster locations, the y-component of the dipole in filtered
maps was correlated with the average temperature anisotropy measured 
from the original {\it unfiltered} map. This average (or monopole) was always negative, 
as expected if the anisotropy is dominated by the TSZ effect. Since in
CMB maps temperature anisotropies are convolved with 
the antenna the correlation was not directly established with
cluster luminosity or mass. But as the TSZ effect scales with X-ray 
luminosity and cluster mass, this correlation was a clear indication 
that both monopole and dipole originated within clusters.

\begin{figure*}
\centering
\includegraphics[width=18cm]{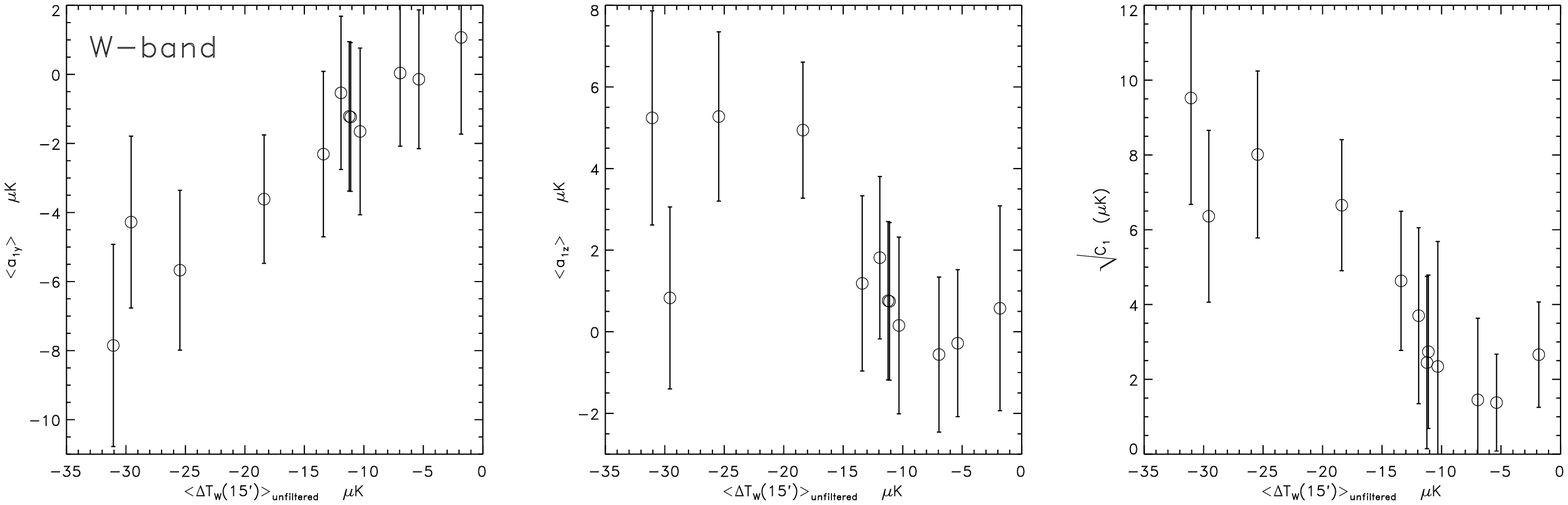}
\caption{\small Results for W-band data per Table 1 of KAEEK 
vs the {\it unfiltered} central monopole. The dipoles were measured
on an aperture of 30arcmin radius and the errors have been computed 
using Method 1 (see Sec.~\ref{sec:3.2}).}
\label{fig:ak4}
\end{figure*}

In Fig.~\ref{fig:ak4} we present the final results from the cluster 
catalog binned by $L_X$ and $z$ per Table 1 of KAEEK. We plot the results 
of the $y,z$ dipole components, evaluated at zero monopole aperture, vs the 
central monopole evaluated from {\it unfiltered} CMB maps. Like in KAEEK
the central monopole was evaluated by averaging over the central $10'$ 
radius; we used only the four W DA's with appropriate angular resolution 
when averaging and we checked that adding the other DA's gives consistent 
results. We show as before we recover statistically significant results 
for the $y$  and  $z$ components; the $x$ 
component is consistent with zero within the errors, but in any event it can 
be derived from the right panel showing the dipole power, $C_1$. As in KAEEK 
the cluster configuration (discussed later for the {\it Planck} analysis) results 
in the value of $a_{1y}$ measured at about $3.3\sigma$ and the value of $a_{1z}$ 
at $\simeq 2.5\sigma$ fully in agreement with Table 1 of KAEEK. 
Fig.~\ref{fig:ak4} shows the same dipole-monopole correlation
than it was found in KAEEK. When one accounts 
for the $L_X$ correlation, the overall ``dark flow" dipole reaches about 
$\simeq 4\sigma$ significance as discussed in AKEKE. In addition, the 
``dark flow" dipole direction coincides with the all-sky dipole direction 
(after correction for the local motion as discussed in Kogut et al. 1993); 
the probability of this happening by chance with the current errors 
is $\sim 10^{-2}$. {\it This argues for the same statistically significant 
signal pointing to the ``dark flow" as in KAEEK}.

\section{The ``dark flow" dipole in Planck data.}
\label{sec:planck}

In March 2013, the {\it Planck} Collaboration released nine {\it Planck} 
Nominal maps from the Low and High Frequency Instruments (LFI and HFI, 
respectively). LFI has measured the CMB sky at frequencies 30, 44 and 
70~GHz, while the HFI has covered the range 100, 143, 217, 353, 545 and 
857~GHz \citep{pr1}\footnote{All {\it Planck} data products have been 
downloaded from the {\it Planck} Legacy Archive found in the following 
web page, 
{\tt http://www.sciops.esa.int/index.php?project=planck\&page=Planck\_Legacy\_Archive}}. 
These maps contain significant foreground contamination due to
synchrotron and free-free emissions at low frequencies 
and thermal dust CO emission and zodiacal light at high 
frequencies. At the HFI frequencies, where zodiacal light 
contribution is more important, we use HFI maps with this 
contaminant removed \citep{pr12}.

In addition, the {\it Planck} Collaboration released four different
foreground-cleaned reconstructions of the CMB temperature anisotropies 
over a large fraction of the sky. These maps were produced 
using the data from nine {\it Planck} channels, without including
any other external dataset, by applying different component separation
techniques. Together with the  foreground-cleaned {\it Planck} Nominal
maps described below, we will analyze the SMICA (Spectral Matching 
Independent Component Analysis) map, constructed from a linear combination in 
harmonic space of the nine single frequency maps of different resolution. 
The weights of each frequency vary with multipole $\ell$. In SMICA the
temperature anisotropy was estimated over 97\% of the sky. The remaining area of
the image shown in Fig.~\ref{fig:planck_maps} is replaced with a 
constrained Gaussian realization \citep{pr12}.
The component analysis used to construct the map does not preserve
the TSZ signal, so it can not be used to test the monopole-dipole
correlation shown in Fig.~\ref{fig:ak4}.

The SMICA map has an angular resolution of 5arcmin, but its harmonic content 
is cut off at $\ell>4000$. The noise has an average rms of $\sim 17\mu$K
with a highly inhomogeneous distribution (see Fig. 15 of \citet{pr1}).
The method under-subtracts thermal dust emission, but at high latitudes, 
in the region where the CMB reconstruction is statistically
robust, residuals are below a few $\mu$K in amplitude.
Compared with other reconstructions using different techniques, SMICA  
produces the map with lowest level of residuals and for this reason 
it will be the one we will be considering here.

\subsection{Cleaning Planck Nominal maps from foreground contributions.}

{\it Planck} Nominal maps contain foregrounds due to diffuse emissions from 
the Galaxy and compact sources. The Galactic foregrounds are the main contaminants 
on large angular scales. The main contributions are synchrotron, free-free and 
the anomalous microwave emission due to spinning dust grains, thermal dust emission 
and emission from CO rotational lines. At small scales, extragalactic foregrounds 
from compact sources and unresolved emission from radio and infrared sources are 
the dominant contribution \citep{pr12}. Foregrounds can be removed through
component separation methods or by reconstructing the foreground fields
and subtracting them. Component separation methods are usually employed
to produce a clean map of CMB temperature anisotropies with very low
foreground contamination at the expense of loosing frequency information
\cite{pr12,bobin1,bobin2}. The Planck Collaboration provides templates to
correct foreground emission at different frequencies and this will
be the approach we will use here. A joint analysis of IRAS and Planck
three highest frequency channels showed that dust varies strongly
on small scales due to dust evolution, extinction and local effects,
particularly in high-contrast molecular regions.
To correct the thermal dust emission we need to take into account
its great variability and to this purpose 
use the {\it Planck} dust model that gives the three parameters 
that define the modified black-body emission law (dust-grains temperature, emissivity 
index and optical depth) at the reference frequency of 353~GHz \citep{pr11}. 
The map of the thermal dust component is given at the same
healpix resolution than the HFI data, $N_{side}=2048$.
To estimate accurately the contribution of this emission at each {\it Planck} 
frequency we evaluate the spectral model in each sky pixel and convolve it
with the passband of each detector. We apply this color correction using
the publicly-available routine {\tt hfi\_color\_correction}. Similarly,
the synchrotron and free-free emissions are cleaned using the data on the
amplitude of those contributions at 30~GHz and a spectral index to scale
it to other frequencies at each pixel on the sky \citep{pes}.
The data is given with a resolution of $N_{side}=256$ and
is integrated with the frequency response at each band using the same 
routine mentioned before to produce the color correction needed
to estimate the flux weighted in each band. The maps of the low-frequency 
and high-frequency foregrounds are subtracted from each frequency map.

The next step of the process is to clean the CO emission. This contribution
is only important for the 100, 217 and 353~GHz channels due to the (1-0), 
(2-1) and (3-2) rotational transition lines. The {\it Planck} collaboration 
has made available three different types of CO correction maps \citep{pr13}. 
Type 1 maps are too noisy to be of use for our purposes, so we use the Type 2 
maps to clean the 100 and 217~GHz channels, the only ones for which the 
correction is available. 

\begin{figure*}
\centering
\includegraphics[width=18cm]{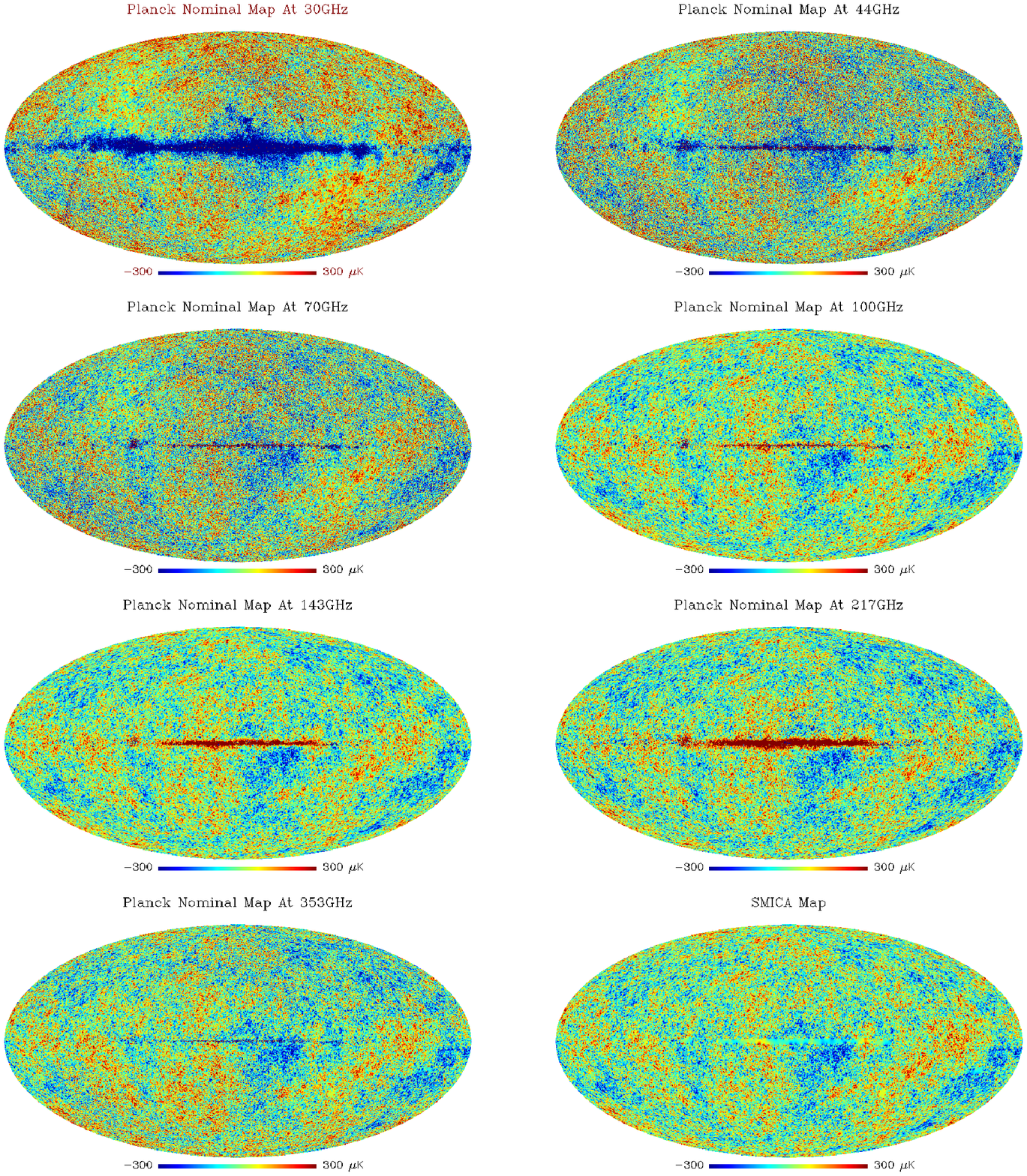}
\hspace*{1cm}
\caption{\small Foreground-cleaned {\it Planck} Nominal maps. From top to bottom
and left to right, maps correspond to channels 30, 44, 70, 100, 143, 217
and 353~GHz; the bottom right plot corresponds to the SMICA CMB map.
}
\label{fig:planck_maps}
\end{figure*}

Our final foreground-cleaned maps from 30 to 353~GHz, together with SMICA,  are shown in 
Fig.~\ref{fig:planck_maps}. The data is plotted  in the range $[-300,300]\mu$K to emphasize
the differences in noise and foreground residuals. We work with a Healpix
resolution of N$_{\rm side}=1024$ \citep{healpix}. In all the maps
there are some residuals of Galactic emission along the Galactic plane but
outside the signal is clearly dominated by CMB fluctuations.
We use the WMAP Kp0 mask to remove the regions around
the Galactic plane and to reduce the contamination due to these
foreground residuals as well as that of point sources.
By using the same mask in {\it Planck} and
{\it WMAP} we use the same fraction of the sky, facilitating the comparison
of the respective results. We did test that the COM-MASK-CMB union mask 
that removes 27\% of the sky and is more adequate to mask point source 
contribution from Planck data, gave the same results than with the Kp0 mask.
Note the stronger similarity of the HFI channels with 
the SMICA map, the reconstruction of the intrinsic CMB anisotropies
made by the {\it Planck} Collaboration, compared with the LFI channels.
The 353~GHz map appears cleaner than the other HFI channels because it was at
this frequency where the {\it Planck} dust model was evaluated. Dust residuals are 
larger at other frequencies due to the uncertainties in the modified black-body 
model or in the determination of the emissivity index. The residual dust
contamination in the Galactic plane diminishes at lower frequencies; at
44 and 70~GHz some residual synchrotron and free-free emission remains.
Also, stripes associated with {\it Planck} scanning strategy 
are clearly seen at 30 and 70~GHz. The satellite preferentially observes the sky
at the ecliptic poles. Since the instrumental noise is higher at 30 and 44~GHz, 
the noise inhomogeneities due to this uneven sampling are clearly noticeable.
We did not consider the maps at 545 and 857~GHz (not shown in 
Fig.~\ref{fig:planck_maps}) due to their stronger foreground contamination
that affected regions of high galactic latitude.

\begin{figure*}
\centering
\includegraphics[width=18cm]{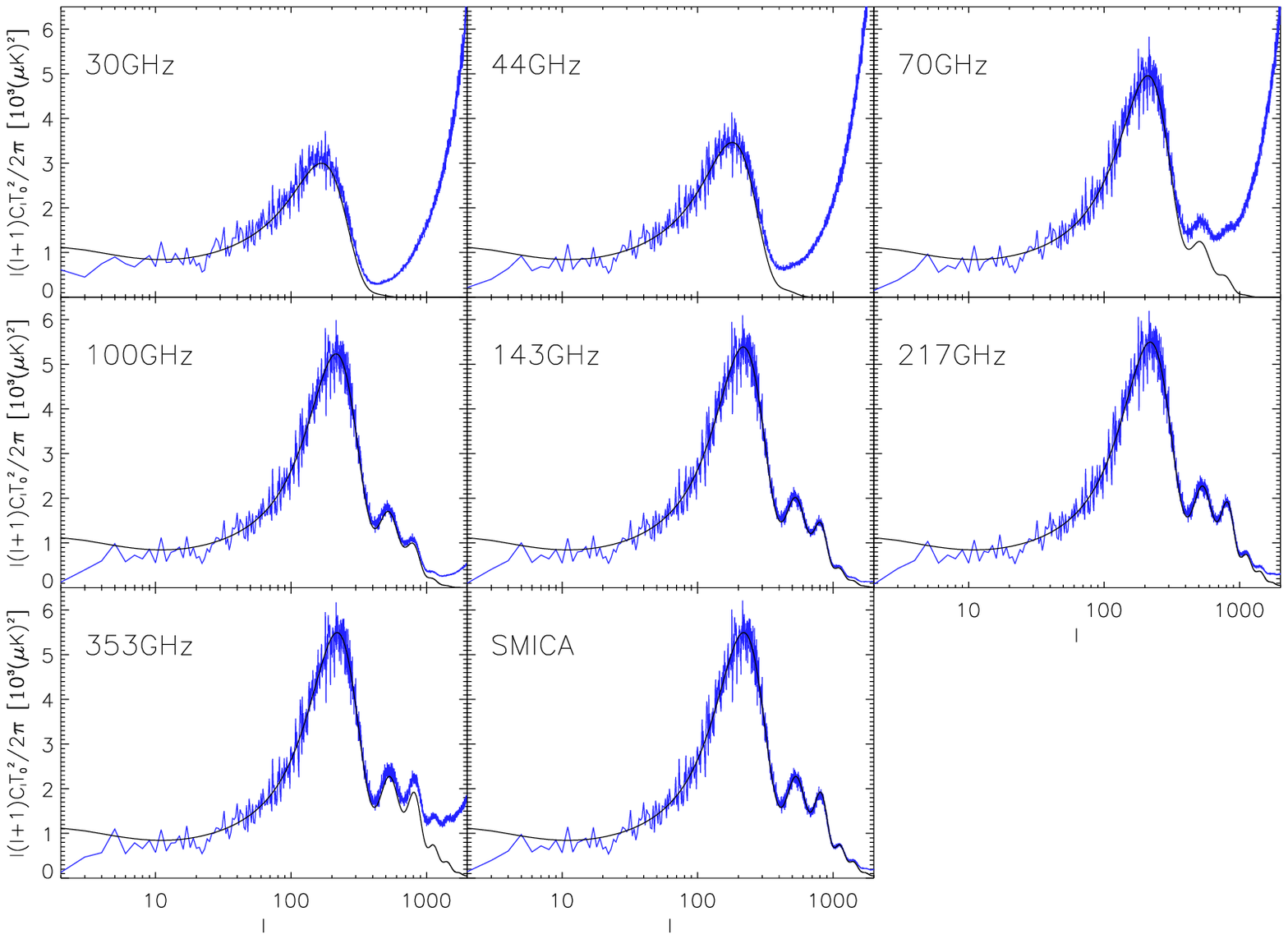}
\caption{\small Radiation power spectra of the maps given in Fig.~\ref{fig:planck_maps}.
The blue and red lines correspond to the first and second frequencies specified 
in the text, respectively. The black lines correspond to the $\Lambda$CDM 
model that best fits the data, multiplied by a gaussian beam at the resolution 
of each channel.
}
\label{fig:planck_cls}
\end{figure*}

\subsection{Planck Data Power Spectrum}

The seven foreground clean maps of Fig.~\ref{fig:planck_maps} have a FWHM of, approximately,
$fwhm=[33^\prime,28^\prime,13^\prime,9.7^\prime,7.3^\prime,5^\prime,5^\prime]$. Their
power spectrum and the theoretical $\Lambda$CDM model, multiplied by the 
antenna beam, are represented in Fig.~\ref{fig:planck_cls} by a
broken blue line and a smooth  solid  black line, respectively. The region around the
galactic plane was masked using the Kp0 mask to remove the
foreground residual contributions near the galactic plane. The theoretical model fits 
the data rather well, with a flat spectrum, noise being the only other 
significant difference. In LFI the noise starts to dominate the intrinsic 
CMB signal at  multipoles $\ell>400-600$,  being much smaller 
in HFI. In none of the spectra there is a deviation
of the theoretical $C_\ell$'s due to foreground residuals or other
artifacts, a reassuring fact that foregrounds have been removed to the levels
required for this project.

\begin{figure*}
\centering
\includegraphics[width=8cm]{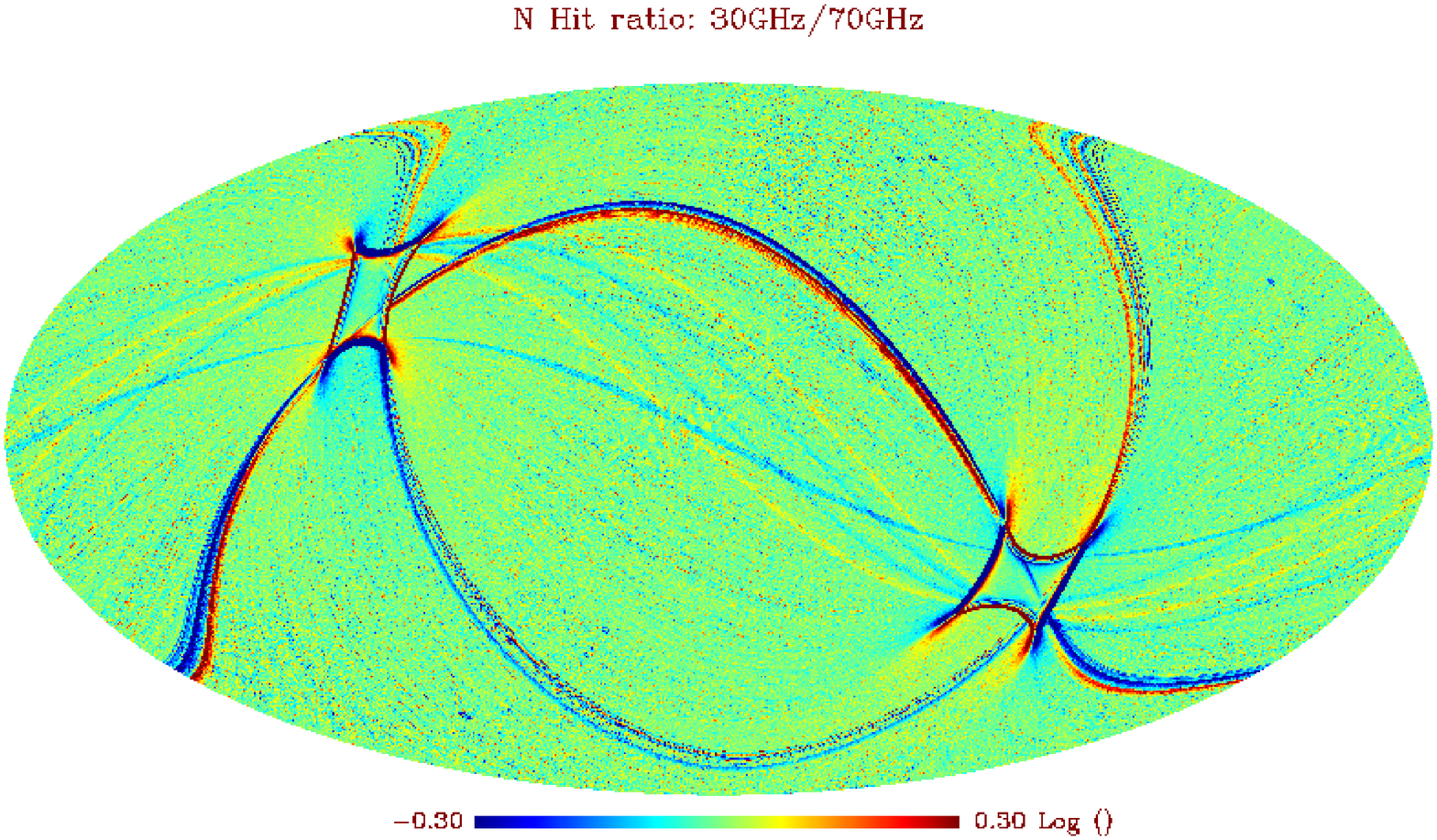}
\hspace*{1cm}
\includegraphics[width=8cm]{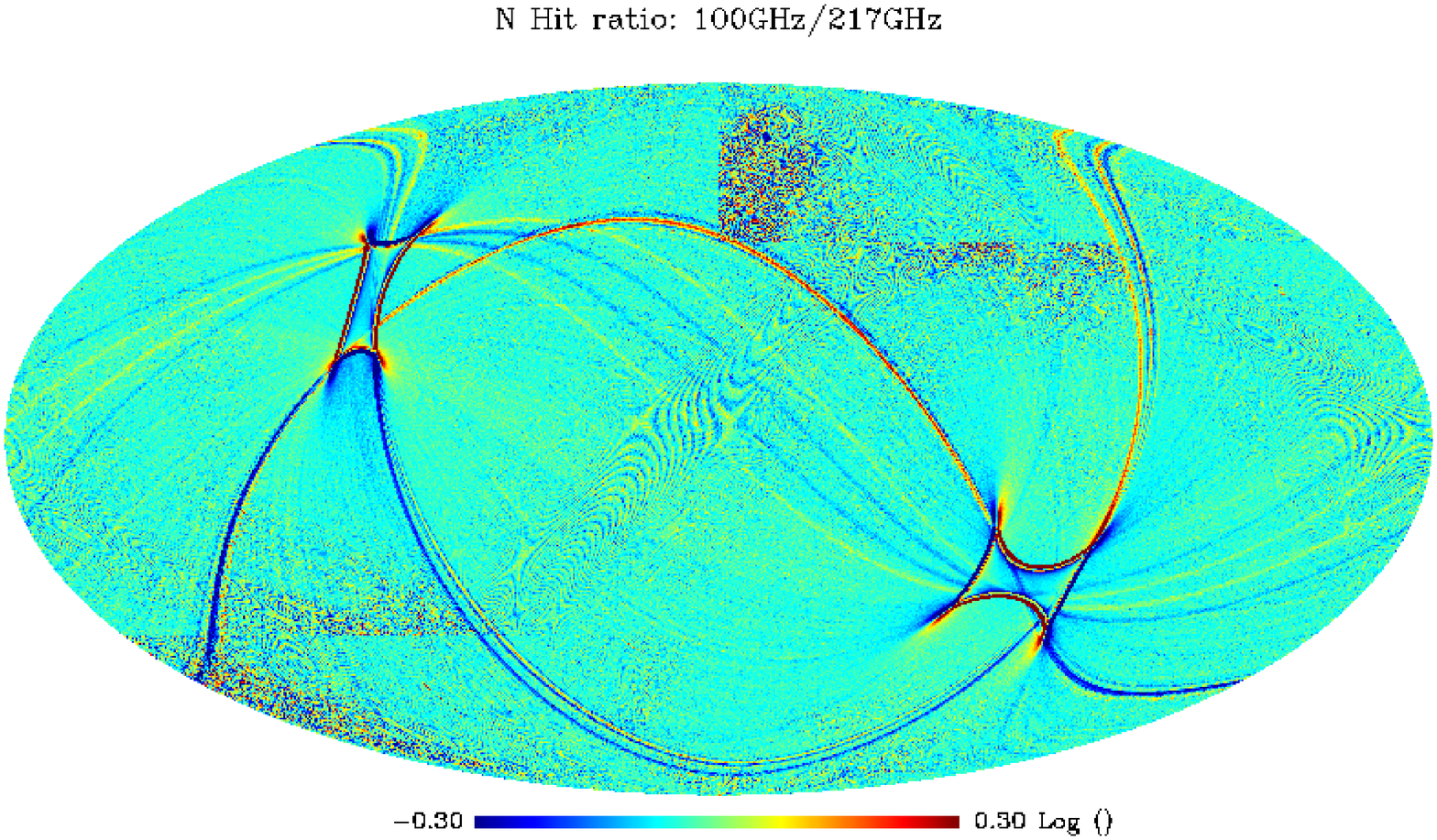}
\caption{\small Ratio of the number of hits between the 30 and 70~GHz (left)
and the 100 and 217~GHz channels (right). A blue stripe corresponds to
number of hits of 70, 217~GHz being larger than those of 30, 100~GHz and
otherwise for a red stripe.
}
\label{fig:stripes}
\end{figure*}

\subsection{Planck Data systematics.}\label{sec4.3}

The {\it Planck} satellite observed the sky at nearly great-circles close
to ecliptic meridians. The times a given position has been scanned (or "hit") 
varies across the sky, giving rise to stripes with a similar pattern at ecliptic
longitudes. The satellite produces one full sky map every 6 months; the initial 
and final position are matched with 6 months difference, when the instruments 
are looking in opposite directions in the solar system. 
Instrument noise, changes in gain, variations on the solar system 
foregrounds -mainly zodiacal light- and other effects contribute to small 
offsets between subsequent scans. As a result,
the data shows stripes at nearly ecliptic meridians, most noticeably
but not exclusively, at the 30 and 70~GHz channels. 

In Fig.~\ref{fig:stripes} we present the ratio of the of the hit maps of
30 and 70~GHz (left) and 100 and 217~GHz (right). Before taking the ratio, 
we divided each hit map by the mean number of 
observations to correct the differences between frequencies. 
Since the detectors at each frequency point to slightly different locations in the sky,
the number of hits at a given location is different.
A blue stripe in Fig.~\ref{fig:stripes} represents a scan where the number of 
hits on the 70, 217~GHz channels is larger than those on 30, 100~GHz and a 
red stripe when it happens otherwise.  If the sky coverage were uniform this ratio would 
be a constant. This figure demonstrates that even if in the foreground-cleaned 
maps of Fig.~\ref{fig:planck_maps} stripes are only seen 
in the LFI maps, they are also present at other frequencies.
It would be important to precisely estimate the effects of the stripes on the noise.
However, without access to the time ordered data or the details of the systematic trends and 
details of which data came from which time we are unable to build a detailed covariance
matrix. If there are residual effects from the stripes in the final data these should show up
as differences when comparing to the WMAP data. The WMAP data has much stronger 
cross-linkages and so its stripes are smaller, at higher spatial 
frequencies and less directional.

Finally, let us remark that the right panel of Fig.~\ref{fig:stripes} 
shows a band of lower number of hits along the ecliptic plane and some 
rather odd features at $(l,b)=(0^0,45^0)$ and at $(270^0,-45^0)$.
We have no explanation about why data, that has been taken along ecliptic 
coordinates, would have been removed along galactic parallels and meridians.
Although we can not estimate the effect of these noise inhomogeneities
on the data, since we compute error bars using the filtered
maps used to compute the dipole, like in {\it WMAP}, 
they will be accounted for in our error bar estimates.

\begin{figure*}
\centering
\includegraphics[width=18cm]{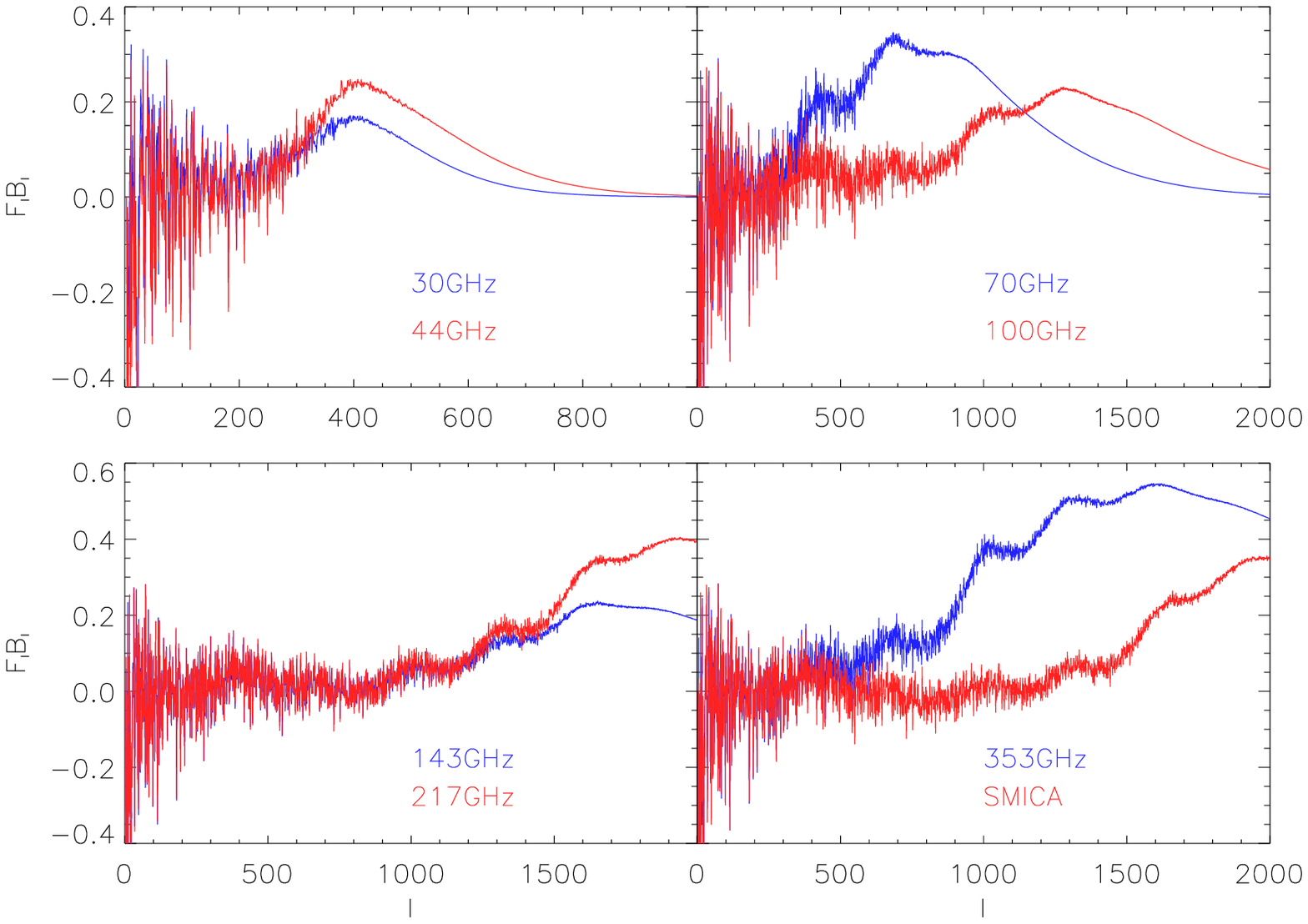}
\caption{\small Filters in $\ell$-space of the seven foreground-cleaned
{\it Planck} and SMICA maps. Lines follow the same convention as in Fig.~\ref{fig:planck_cls}.
}
\label{fig:planck_wf}
\end{figure*}

\begin{figure*}
\centering
\includegraphics[width=18cm]{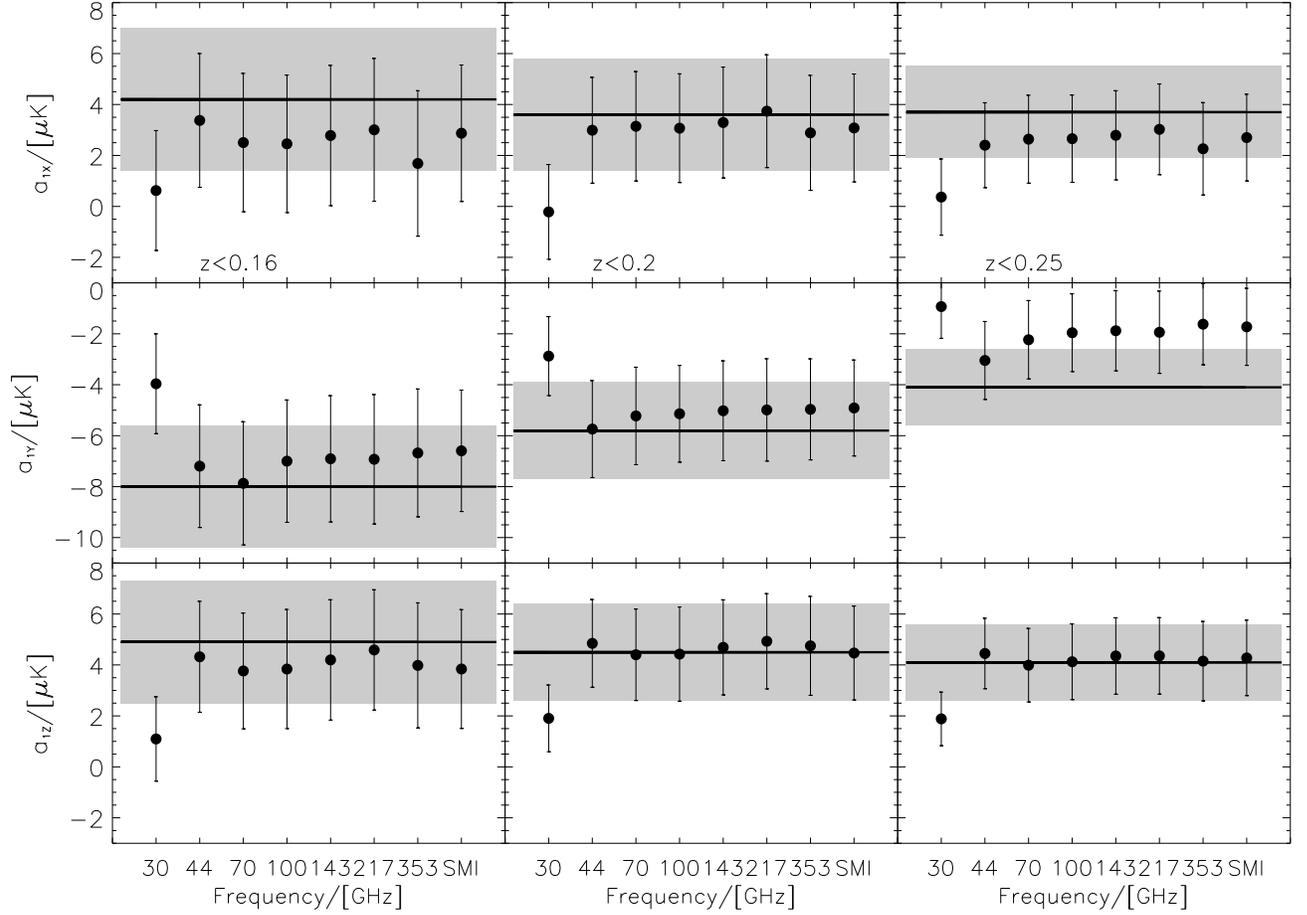}
\caption{\small Comparison of {\it WMAP} 5 yr and {\it Planck} 1 yr dipoles. 
{\it Planck} dipoles are represented by filled circles with the corresponding 
error bars computed using Method 1 (see Sec~\ref{sec:3.2}).
Solid lines and shaded areas correspond to the measured dipoles 
in WMAP 5 yr data for the same cluster configurations, as in Fig.~\ref{fig:ak2}.
}
\label{fig:planck_results1}
\end{figure*}

\begin{figure*}
\centering
\includegraphics[width=18cm]{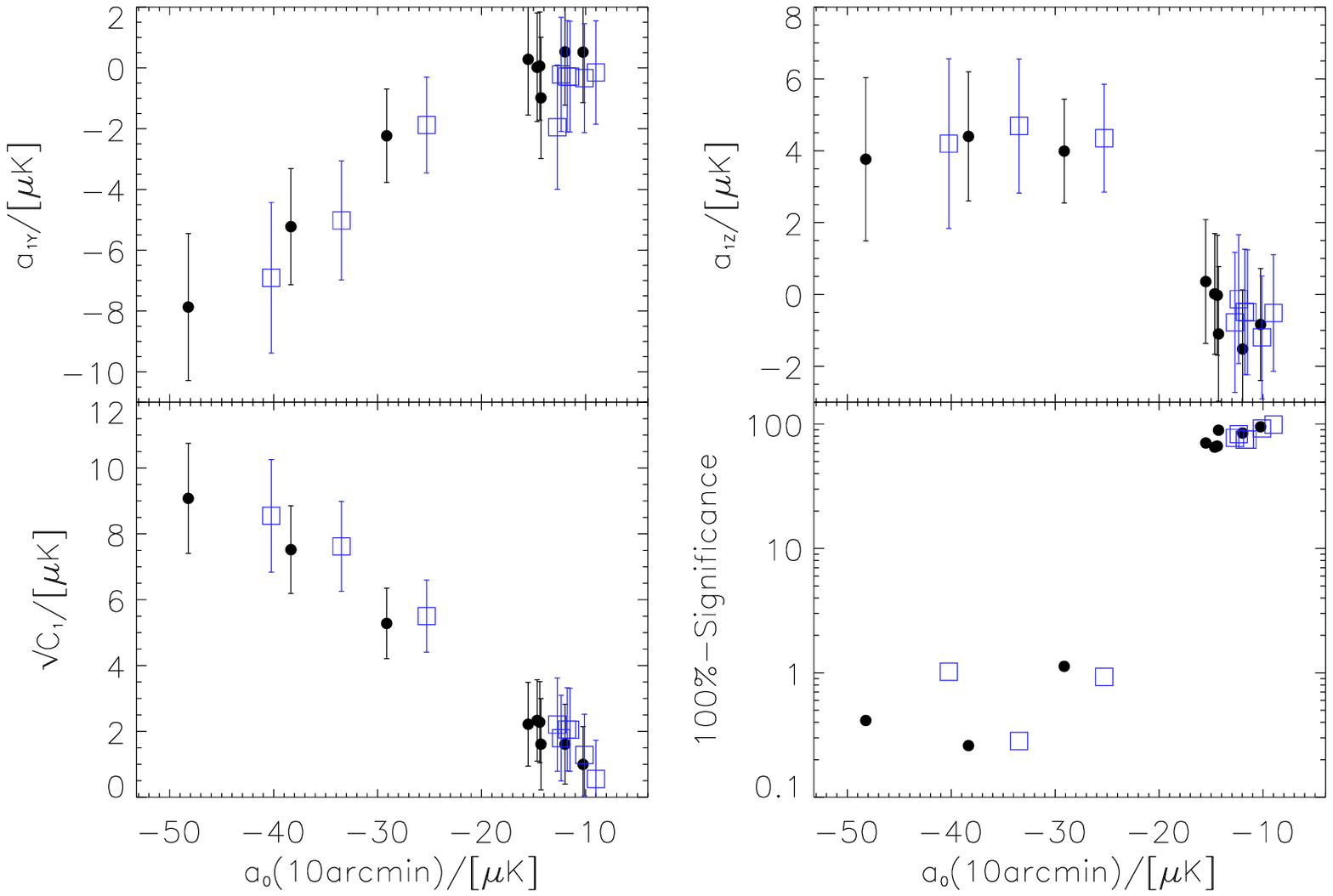}
\caption{\small Dipoles for all the cluster configurations measured 
on apertures of 25' vs the unfiltered monopole on 10' and their statistical
significance. Error bars have been computed using Method 1. Only two channels are shown:
70~GHz (black solid circles) and 143~GHz (blue squares).
}
\label{fig:planck_results2}
\end{figure*}

\subsection{Results}

{\it WMAP} and {\it Planck} scanning strategies are very different, so comparison 
between both data sets is important to isolate systematics.
To facilitate the comparison, we filter the CMB signal
and compute the dipole at the cluster location following the same
steps as in {\it WMAP}. We use the same mask in both datasets, the Kp0 mask.
As the input theoretical model we use the 
$\Lambda$CDM radiation power spectrum with the {\it Planck} measured parameters. 
The filters of the seven {\it Planck} Nominal maps and the SMICA map
are presented in Fig.~\ref{fig:planck_wf}. Blue and red lines corresponds
to the first and second frequency indicated in each panel. Due to the
higher noise levels in LFI channels, at $\ell\ge 500$ the filter is close
to unity while it oscillates around zero up to $\ell\simeq 10^3$ for the
HFI channels and the SMICA map. Compared with 
the {\it WMAP} filters shown in Fig.~\ref{fig:ak1}, the filters of 
the two lowest LFI frequencies are similar to 
those of {\it WMAP} but the other filters are very different
due to differences in resolution and instrument noise. This behaviour
can be explained by the functional form of our filter, $F_\ell=
(C_\ell^{sky} - C^{th}_\ell B^2_\ell)/C^{sky}_\ell$;
when the noise dominates $F_\ell \simeq 1$ and the graph of $F_\ell B_\ell$
behaves like $B_\ell$. This is clearly seen in WMAP and the two lowest LFI 
frequencies at $\ell\ge 400$. When the noise is negligible, as in SMICA, 
then $C_\ell^{sky}$
differs from $C^{th}_\ell B^2_\ell$ due to cosmic variance and $F_\ell$
oscillates around zero till the noise starts to dominate. For the 
maps with the lowest noise levels, this happens at higher $\ell$'s.
The exponential cut-off due to the antenna beam occurs at $\ell\sim 1500-2000$, 
almost outside the multipole range shown in Fig.~\ref{fig:planck_wf},
giving the graph an overall different aspect.
Therefore, {with {\it Planck} data} we can test the effect of the 
filter on our results more than it could do with {\it WMAP}.

In Fig.~\ref{fig:planck_results1} we present the dipoles
of the same three cluster configurations of Fig.~\ref{fig:ak2}. Error
bars were computed using Method 1: We generated 1,000 templates of 800 
disks randomly placed on the sky and computed the dipoles
on the filtered maps of those random templates. For each template we took
subsets of 100, 200 and 400 clusters and verified that 
$\sigma_{1m}\propto N_{cl}^{-1/2}$. Like in {\it WMAP}, error bars on the 
dipole are driven by the error on the monopole and are approximately
given by eq.~\ref{eq:error}.

The results presented in Fig.~\ref{fig:planck_results1} show a remarkable 
consistency among themselves and with those of {\it WMAP} 5yr data. 
The measured dipoles are {\it independent of frequency}
(with the exception of 30~GHz) {\it and the filter} used. The filters,
shown in Fig.~\ref{fig:planck_wf}, have different structure in $\ell$-space 
since have been designed to remove the intrinsic CMB anisotropies attending 
to the specifics of each particular data set. Only the 
theoretical model $C_\ell^{th}$ is common to all filters. The consistency of 
the measured dipoles shows that our results are neither generated by artifacts
introduced by our pipeline nor are dominated by systematics present on the data.
The spectral distribution confirms our earlier findings with WMAP:
The dipole can not be due to the TSZ effect or from a systematic
associated with foreground residuals in the data as 
the measured dipole remains constant at all frequencies except for
the offset at 30~GHz, which is most stripe-dominated.
The dipole is clearly different of what it would be if it was due
to TSZ effect as suggested by \citet{omcp}. 
In this case, the dipole had to be zero at 217~GHz and
have the opposite sign at 353~GHz, none of which is observed.
It is also different from what it would be if it was due to
foreground residuals that correlated with cluster properties.
All known foregrounds vary with frequency, contrary to what
is shown in Fig.~\ref{fig:planck_results1}, where the dipoles
between 40 to 353~GHz remain constant, independent of frequency.
Only the dipoles measured in the filtered 30~GHz map appear to be
systematically different (at $\sim 1\sigma$ level) and closer to zero than 
those of all the other maps, including the SMICA map. At the other
{\it Planck} frequencies the dipoles are slightly offset compared
with the values measured in WMAP. For example, at $z\le 0.25$ $a_{1Y}$ 
is systematically 
above the KAEEK value at more than $1\sigma$. We will discuss later the possible 
reasons for systematic differences between {\it WMAP} and {\it Planck} results.

As in Fig.~\ref{fig:ak4}, the dipoles measured in {\it Planck} 
data show a clear correlation with the TSZ monopole in the unfiltered map, 
both in the LFI and the HFI channels.
In Fig.~\ref{fig:planck_results2} we show the 
dipole components, $a_{1Y}$ and $a_{1Z}$, the dipole modulus
and the statistical significance for the cluster configurations
selected according to redshift $z\le 0.16, 0.2, 0.25$ 
and X-ray luminosities (in units of $10^{44}$erg/s) in the
range $L_X<1$, $L_X=[1-2]$ and $L_X>2$. The dipoles
are plotted vs the monopole $a_0$ measured over a solid aperture
of radius 10arcmin in the original (unfiltered) foreground-cleaned {\it Planck}
Nominal maps. For simplicity we only show two frequencies:
70~GHz (solid black circles) and 143~GHz (blue squares). 
The statistical significance has been computed by
generating $10^5$ random dipoles with zero mean and rms
dispersion the uncertainty in each dipole component for each cluster
configuration and finding the fraction of random dipoles with amplitude
larger than the measured value. The statistical significance 
exceeds 99\% in the three most significant bins,
the three bins with the brightest clusters, $L_X>2\times 10^{44}$erg/s.

First, notice that at 70 GHz monopoles are larger than those at
143GHz, but the ratio is smaller than $\sim 1.7$, i.e., it is
smaller than the ratio of the TSZ amplitude at 70GHz to 143GHz.
The 70GHz channel has lower resolution (see Sec~4.2) and dilutes cluster 
anisotropies more than 143GHz.  Second, the largest monopoles 
and dipoles corresponds to the most luminous 130 clusters with $z<0.16$ but
the highest significance corresponds to the 208 clusters with $z<0.2$.
The second bin has a larger number of clusters and the dipole components 
are measured with a slightly better signal-to-noise ratio. This small difference 
results in great variations on the statistical significance since we are exploring
the tail of the distribution.

\subsection{Comparison of WMAP and Planck dipoles.}

\begin{figure*}
\centering
\includegraphics[width=18cm]{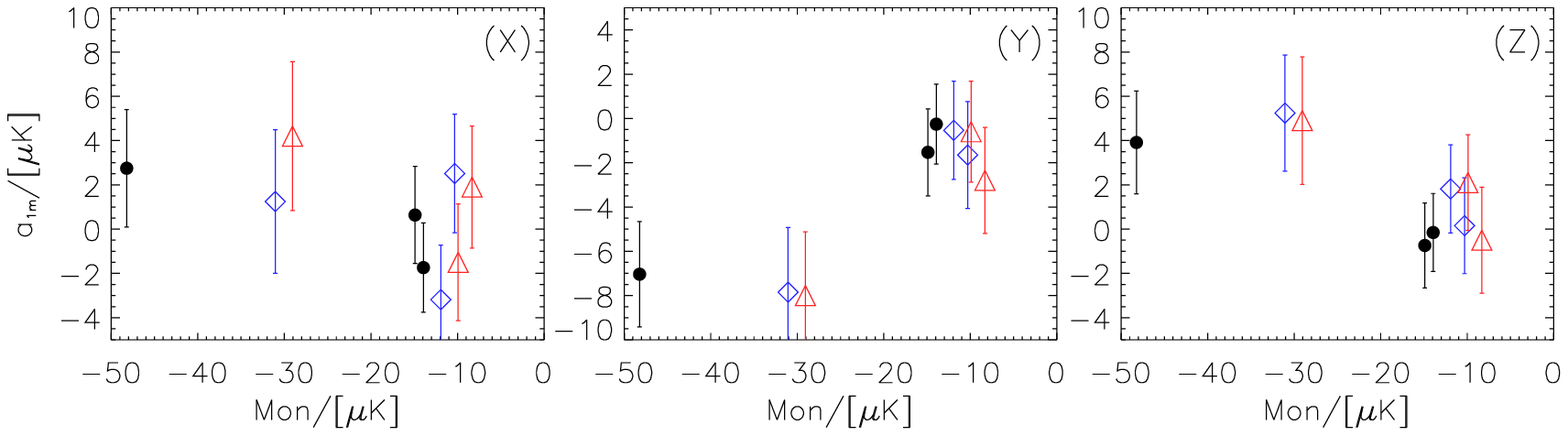}
\vspace*{-7.5cm}
\caption{\small Comparison of {\it WMAP} and {\it Planck} dipoles for
the three luminosity bins with of clusters with redshift $z\le 0.16$.
Triangles (red), diamonds (blue) and solid circles correspond to
{\it WMAP} W-band 5, 9 yr data and {\it Planck} data, respectively.
\bigskip
}
\label{fig:wmap59_planck}
\end{figure*}

The dipoles measured at the different {\it Planck} frequencies display remarkable
consistency, except for 30~GHz, and exhibit a strong correlation with the TSZ monopole.
These results are consistent with the dipoles measured previously in {\it WMAP}.
In Fig.~\ref{fig:wmap59_planck} we compare the three components
of the dipole for the three X-ray luminosity bins of clusters with
redshift $z\le 0.16$ measured in {\it WMAP} 5, 9yr and {\it Planck} at
100GHz, represented by triangles (red), diamonds (blue) and solid circles
(black), respectively. In the X-axis, the 5 yr data {\it is shifted by two units} to 
avoid over-plotting data. The results from {\it WMAP} 3 and 7 yr
data are also consistent with those plotted in the figure but are 
not shown to avoid overcrowding. The monopoles are computed on apertures 
of 10arcmin in {\it unfiltered} maps while dipoles are computed over apertures 
of 25arcmin radii, which correspond to the {\it WMAP} zero monopole aperture.
Although the {\it WMAP} W-band frequency of observation is 94GHz, not 
far from the 100GHz channel shown in the figure,  its monopoles
are significantly smaller, particularly for the bin containing  
the most luminous clusters. The difference is due to {\it WMAP} having 
lower angular resolution ($\sim 12'$) than {\it Planck} ($\sim 5'$) at 
those frequencies and, consequently, the TSZ cluster anisotropy is more diluted.

\begin{figure*}
\centering
\includegraphics[width=18cm]{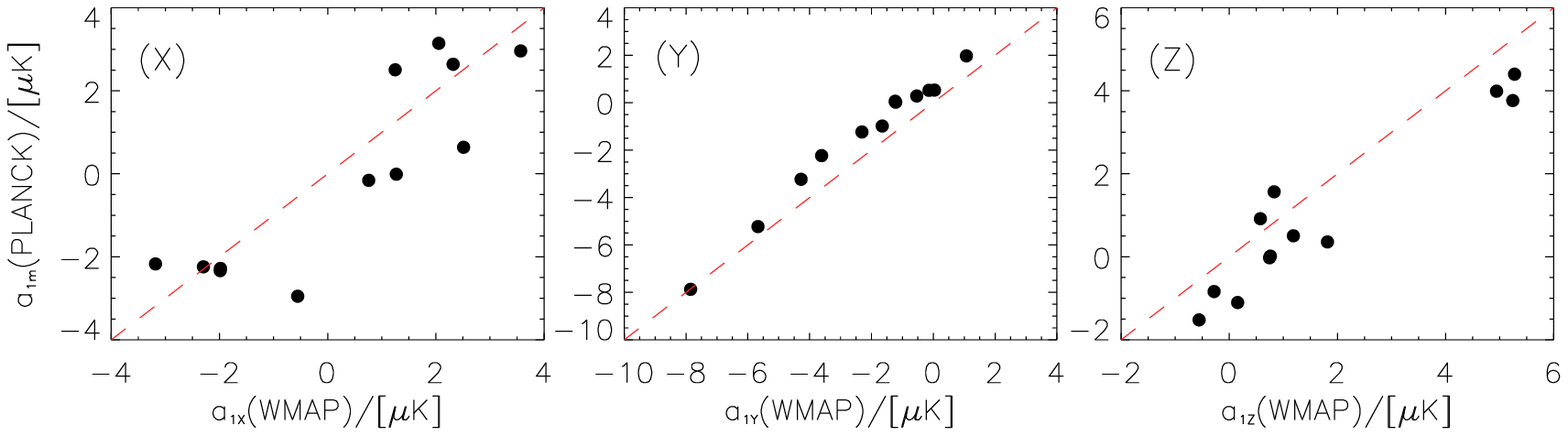}
\vspace*{-7cm}
\caption{\small Comparison of the three dipole components measured in {\it WMAP} 9yr
W-band  data with those of {\it Planck} at 70~GHz for the same cluster subsamples. 
The dashed red line corresponds to the dipoles been equal.
For clarity, error bars are not shown. The differences between both data
sets are always smaller than $1\sigma$ for all configurations.
\bigskip
}
\label{fig:wmap_planck}
\end{figure*}

While {\it WMAP} and {\it Planck} data are consistent, the data
show some small but systematic differences. In Fig.~\ref{fig:wmap_planck} we compare 
the dipoles of {\it WMAP} 9-yr data, averaged over the 4 DA's of the W band with 
the dipoles measured in {\it Planck} 70~GHz map, of similar angular resolution.
Only 11 points (out of the 12 cluster subsamples of KAEEK) are seen since 
two values merge on the plot, as indicated in Sec.~\ref{sec:wmap}.
For reference, the red dashed line shows the dipoles having the same
value in both data sets. For clarity we do not show the error bars here;
for any cluster subsample the dipoles of either satellite differ by less
than one standard deviation. Fig.~\ref{fig:wmap_planck} shows that the $a_{1x}$'s
components are randomly distributed above and below the red dashed line, 
the $a_{1y}$'s measured in {\it Planck} are systematically
smaller than those of {\it WMAP} and the distribution of 
the $a_{1z}$'s is indifferent.

Although the discrepancies in Fig.~\ref{fig:wmap_planck} are not yet relevant, 
the systematic offset in the $y$-component of the dipole or the discrepant
results of the 30~GHz channel could be the result
of systematics present in the {\it Planck} data.
We have already noted in Figs~\ref{fig:planck_maps} and \ref{fig:stripes} 
that foreground-cleaned {\it Planck} Nominal maps contain very strong non-gaussian 
features. Even if stripes correspond to
differences in the number of observations (or weights) in the data,
they are not erased but are rather enhanced by filtering. 
The filter depends on the noise, $(1/f)$-features 
could introduce some effect, mostly in the 30~GHz map, the channel
with the largest intrinsic noise of all the {\it Planck} frequencies.
There are other effects that could be more pernicious.  The low weight 
stripes are often there because over most of the sky there are 2 distinct
sets of observations 6 months apart while at the beginning or end of the 
period there is only one set of observations.  This has several effects, first
because there is only one set of data, there is a single correction for
long term drifts so whatever effects are there one has fewer data sets to 
average over.  This increases the systematic effects according to the 
number of data sets which is a small number. Secondly there are fewer
data to check these long term drifts against other data. Finally in the middle of
a data set one can interpolate, while at the end one must necessarily
extrapolate which is intrinsically more uncertain. 
These effects lead to higher noise at low frequencies. The pattern of the 
{\it Planck} observations puts the low temporal frequencies at low spatial 
frequencies. As can be noted from Fig.~11 of \citet{pr2}
systematic effects have higher than proportional noise at low frequencies
and this effect would be largest at 30~GHz, where the noise is higher,
than at other channels and could be the underlying reason why the
dipole has not been equally preserved by the filter than in WMAP.
As indicated in Sec~\ref{sec4.3}, understanding the effect of 
stripes would require to analyze the time ordered data and goes
beyond the scope of this paper.

\subsection{Comparison with Planck earlier results}

Our results differ markedly from an earlier analysis of {\it Planck} data
using the Internal Linear Combination map (\citet{pir13}, hereafter PIR-13), 
a foreground clean map similar to SMICA constructed to measure 
the KSZ effect. The TSZ contribution in their map was removed to less than a 
few percent of its original value. The {\it Planck} Collaboration claim not
to have found any detection of a bulk flow as measured in any comoving sphere extending 
to the maximum redshift covered by their cluster sample. In fact, they found a
dipole for their full cluster sample (see Fig.~10 of PIR-13) similar to ours but
overestimated their error bars, diluting the statistical significance of
their measurement. The Planck Collaboration used two flawed methods to 
compute errors. 1) 
They rotated the cluster template around the z-axis; this method underestimates
the error on the z-component and overestimates it in the  x and y-components, given an
overall increment on the error of modulus (see Atrio-Barandela (2013),
Table~1 and Figs.~2 and 3). 2) They computed the errors measuring dipoles with
the actual distribution of clusters over simulated CMB skies, but 
their simulations did not mimic the data accurately enough. 
In the real sky, filtering leaks power from high galactic latitude to the Galactic
plane. In their simulated maps the Planck Collaboration did not apply any galactic
mask, preventing the power leakage to the plane of the Galaxy. As a result, their simulated
maps contain higher power than the actual sky (see Atrio-Barandela (2013) Figs.~4 
and 5), again overestimating their errors by a similar amount than in their 
rotation method\footnote{This information was passed on by 
FAB when PIR-13 was being written.} (see Atrio-Barandela 2013 for a full discussion).
In addition, they did not find a larger dipole for their 200 most massive clusters.
While some differences may arise from the differences in cluster samples,
the lack of correlation between dipole and monopole is probably due (A) to 
their binning and (B) to having eliminated the TSZ component from their 
map, a component which we have shown not to have an effect on the measured dipole
(see Fig.~\ref{fig:planck_results1}). In Figs.~\ref{fig:ak4} and 
\ref{fig:planck_results2} we have demonstrated that the largest dipole originates
from clusters with $L_X>2\times 10^{44}$erg/s and $z<0.16$. Adding clusters
with higher redshift in the same luminosity bin reduces both the monopole and 
dipole, consistent with clusters being more diluted by the antenna. If their
subsample of massive clusters is, on average, at higher redshifts,
then their dipole would be smaller than the values we have found. 
To verify the dipole-monopole correlation they ought to
have measured the TSZ from foreground clean maps at different frequencies,
to check if their most massive clusters produced the largest TSZ monopole or not. 

\section{Conclusions} 
\label{sec:conclusions}

We have computed the dipole at the cluster locations using the same techniques
for {\it WMAP} and {\it Planck}. We find a ``dark flow'' signal which correlates 
with X-ray properties, and is therefore likely related to cluster gas, and not 
to the primary CMB, foregrounds or noise. The results are in excellent agreement 
with our earlier findings and are consistent both in {\it WMAP} 9 yr and in 
{\it Planck} 1 yr.  Those instruments used
different scanning strategies, which resulted in different systematics and, while
small differences remain, the close agreement is reassuring of the real 
nature of the dipole signal. Specifically we found that:
\begin{itemize}
\item{} The dipole at the zero monopole aperture remains at 
cluster positions at the same level as in KAEEK.
\item{} The dipole at cluster positions correlates with the TSZ monopole,
a proxy for X-ray luminosity. 
\item{} The signal is consistent among the different multi-year {\it WMAP} 
integration filters and with all {\it Planck} frequencies, except for a small, 
typically $\sim 1\sigma$, offset at 30~GHz.
\item{} The noise of the measurement in our filtered maps is in good agreement with 
the analytical and numerical theory developed in AKEKE and summarized here.
\item{} The overall statistical significance of the dipole signal in {\it WMAP}
is similar to that found in KAEEK, and is larger for {\it Planck} than for {\it WMAP}.
\item{} Within the uncertainties the signal points in the direction of the all-sky CMB dipole.
\item{} If one accepts the KSZ interpretation of the detected statistically significant 
signal the equivalent velocity is $\sim 600-1,000$km/s, within the systematic 
and statistical calibration uncertainties discussed by Kashlinsky et al. (2009), 
KAEEK and Atrio-Barandela et al. (2012).

\end{itemize}

While we deliberately avoid interpretation here, we note that 
the measurements are consistent with the ``dark flow" proposition (KABKE), namely 
the existence of a primordial CMB dipole of non-kinematic origin, which then
presents itself as an effective motion across the entire cosmological horizon. 
No other alternative interpretation of the measured signal has been advanced
although it would be of scientific interest. Instead, 
the debate concentrated along the lines of trying alternative filtering schemes, 
which may erase the signal (Atrio-Barandela et al. 2012). Indeed, an all-sky 
filtering cannot imprint a dipole exclusively at cluster positions, which would 
in addition correlate with cluster X-ray luminosity, but given the still 
limited significance of the measurement of about $(3-4)\sigma$, other 
filtering schemes can reduce the measurement below being statistically 
significant (see Fig. 13 of \cite{omcp}, where such alternative filtering 
schemes start picking up the KSZ signal at velocities exceeding 4,000-6,000 km/sec).

If the ``dark flow'' corresponds to a large scale motion it is of interest 
to compare with peculiar velocities derived using other methods. First, \citet{pr27} 
have measured the aberration of the CMB temperature fluctuations due to our
local motion, constraining the amplitude of large scale flows in the direction
of the solar motion, i.e., constraining the motion of the Local Group
projected in that direction but not the full vector. Velocity
estimates relying on distance indicators are affected by their uncertainties.
For instance, Watkins \& Feldman (2014) argue that their previous results 
overestimated the flow due to their distances being underestimated
by 10\%. Probes of the velocity field on scales $\ge 100h^{-1}$Mpc
depend on the value of the Hubble constant and the current discrepancies
between local measurements and the {\it Planck} value makes these measurements
even more uncertain.  Supernovae Type Ia have also been used to measure velocities. 
Turnbull et al. (2013) find that their sample does not show large-scale 
bulk flow.  However, Wiltshire et al. (2013) argue for the opposite finding 
that the Hubble expansion exhibits considerably more variance in the rest-frame of 
the CMB dipole that in the inertial frame of the Local Group. 
The cosmic radio dipole is also peculiar. It has an
amplitude larger than expected from a purely kinematic effect and a significant
contribution to this excess could come from a local void or similar structure
\cite{dominik}. Future work, 
including by our team, with an expanded cluster catalog now in an advanced stage
of preparation, should shed more light on the existence of the ``dark 
flow''. 

\begin{acknowledgements}
FAB acknowledges financial support from the Spanish
Ministerio de Educaci\'on y Ciencia (grant FIS2012-30926). 
We thank R. G\'enova-Santos for providing the foreground
cleaned maps used in this work.
\end{acknowledgements}

\pagestyle{plain}

\end{document}